\documentclass[aps,pre,twocolumn,groupedaddress,showkeys]{revtex4}
\usepackage{graphicx}

\begin{document}

%\preprint{}

\title{Coarse-Grained Picture for Controlling Quantum Chaos}


\author{Toshiya TAKAMI}
%\thanks{}
\affiliation{Institute for Molecular Science, Okazaki 444-8585, Japan}
\author{Hiroshi FUJISAKI}
\affiliation{Department of Chemistry, Boston University,
590 Commonwealth Ave.,
Boston, Massachusetts, 02215, USA}
\author{Takayuki MIYADERA}
\affiliation{Department of Information Sciences, Tokyo University of Science,
Noda City, 278-8510, Japan}


\date{\today}

\begin{abstract}
We propose a coarse-grained picture to analyze control
problems for quantum chaos systems.  Using optimal control
theory, we first show that almost perfect control is achieved
for random matrix systems and a quantum kicked rotor.  Second,
under the assumption that the controlled dynamics is well
described by a Rabi-type oscillaion between unperturbed
states, we derive an analytic expression for the optimal
field. Finally we numerically confirm that the analytic field
can steer an initial state to a target state in random matrix
systems.
\end{abstract}


\maketitle

%\section{}
%\subsection{}
%\subsubsection{}

\section{Introduction}

Controlling quantum systems 
is one of hot topics
in physics and chemistry as 
illustrated in the fields of 
quantum information processings \cite{NC00,RR96,TV02} 
and laser control of atomic and molecular processes \cite{RZ00}.
As for the latter, there have been devised various control schemes: 
A $\pi$-pulse is a simple example to 
induce a transition between two eigenstates \cite{AE87}.
As a generalization of the $\pi$ pulse or adiabatic rapid passage \cite{MGHGW94}, 
we can utilize the nonadiabatic transitions induced by laser fields \cite{TN98}.
For more than three level systems, 
STIRAP scheme uses a counterintuitive 
pulse sequence to achieve 
a perfect population transfer between two eigenstates \cite{BTS98}.
When more than two electronic states are involved in the controlled system, 
we can use a pulse-timing control (Tannor-Rice) scheme 
to selectively break a chemical bond on a desired potential surface 
by using a pump and dump pulses with an appropriate time interval \cite{TR85}.
When the controlled system  has more than two pathways
from an initial state to a target state,
quantum mechanical interference between them
can be utilized to modify the ratio of products,
which is called coherent control (Shapiro-Brumer) scheme \cite{SB03}.

These control schemes are very effective for a certain class of processes 
but are not versatile and ineffective for, e.g., multi-level-multi-level 
transitions we shall consider in this paper. 
There exist several mathematical studies 
which investigate controllability
of general quantum-mechanical systems \cite{HTC83,PDR88}.
The theorem of controllability says that
quantum mechanical systems with a discrete spectrum
under certain conditions 
have
complete controllability in the sense that
an initial state can be guided to a chosen target state after some time.
Although the theorem guarantees the existence of optimal fields,
it does not tell us how to construct 
such a field for a given problem.

One of the method to practically design an optimal field
is optimal control theory (OCT) \cite{PDR88,ZBR98} or genetic 
algorithms \cite{JR92,RZ00}.
We focus on the former in this paper as a theoretical vehicle.
The equations derived from OCT are highly nonlinear (and 
coupled), so we must solve them using some iterative 
procedures.
There are known some effective algorithms to carry out this procedure 
numerically, however, the field thus obtained is so complicated
that it is difficult to analyze the results: 
What kinds of dynamical processes
are involved in the controlled dynamics?
In addition, the cost of the computation becomes larger
if we want to apply OCT to realistic problems with many degrees of freedom.
Several efforts have been paid to reduce computational costs;
Zhu and Rabitz \cite{ZR99} have introduced a non-iterative 
algorithm for the optimal field. 

On the other hand,
we know that some chemical reaction systems, especially when highly excited,
exhibit quantum chaotic features \cite{Gutzwiller90}, i.e.,
statistical properties of eigen-energies and eigen-vectors 
are very similar to those of random matrix systems \cite{Haake01}.
We call such systems {\it quantum chaos systems} in short.
It has been also studied how these quantum chaos systems behave 
under some external parameters \cite{GRMN90,Takami91,ZD93}.
These statistical properties of quantum chaos systems
stem from multi-level-multi-level interactions of eigenstates,
which are related to the existence of many avoided crossings \cite{Takami92}.
Hence it is necessary to consider the interaction between many eigenstates
when we study dynamics in such a system.
Furthermore, 
if our purpose is to control a Gaussian 
wavepacket in a quantum chaos system, 
the process also becomes a multi-level-multi-level 
transition because a Gaussian wavepacket 
in such a system contains many eigenstates.
These are our motivations why we treat multi-level-multi-level transitions 
and want to control them.

This paper is organized as follows.
In Sec.~\ref{sec:oct-numerical},
we show how quantum chaos systems
can be controlled under the optimal fields 
obtained by OCT.
The examples are a random matrix system
and a quantum kicked rotor. 
(The former is considered as a strong-chaos-limit case 
and the latter as mixed regular-chaotic cases.)
In Sec.~\ref{sec:cg-picture},
a ``coarse-grained'' Rabi state is introduced
to analyze the controlled dynamics in quantum chaos systems.
We numerically obtain a smooth transition between time-dependent 
states, which justifies 
the use of such a picture.
In Sec.~\ref{sec:analytic},
we derive an analytic expression for the optimal field
under the assumption of the CG Rabi state, 
and numerically show that the field 
can really steer an initial state to a target state 
in random matrix systems.
Finally, we summarize the paper
and discuss further aspects of controlling quantum chaos.

\section{Optimal Control of Quantum Chaos\label{sec:oct-numerical}}

We study optimal control problems of quantum chaos systems.
Our goal of control is to obtain
an optimal field $\varepsilon(t)$ which guides a quantum chaos system
from an initial state $|\varphi_i\rangle$ at $t=0$ to
a given target state $|\varphi_f\rangle$ at some specific time $t=T$.
One such method
is optimal control theory (OCT), which
has been successfully applied to atomic and molecular systems \cite{RZ00}.

OCT is usually formulated as a variational problem under constraints
as follows:
We start from the following functional used by Zhu-Botina-Rabitz \cite{ZBR98} 
\begin{eqnarray}
\label{eqn:functional}
&&J=J_0-\alpha\int_0^T [\varepsilon(t)]^2 dt\nonumber\\
&&-2{\rm Re}\left[
  \langle\phi(T)|\varphi_f\rangle
  \int_0^T\!\!\langle\chi(t)|
    {\partial\over\partial t}-{H[\varepsilon(t)]\over i\hbar}|\phi(t)\rangle dt
\right].
\end{eqnarray}
The first term in the right-hand side is the squared absolute value
of the final overlap,
\begin{equation}
  J_0=|\langle\phi(T)|\varphi_f\rangle|^2.
\end{equation}
The second term is the penalty term
with respect to an amplitude of the external field $\varepsilon(t)$.
The factor $\langle\phi(T)|\varphi_f\rangle$ in the last term is introduced
to decouple the conditions for the state $|\phi(t)\rangle$ and
the inversely-evolving state $|\chi(t)\rangle$, 
which both evolve under the Hamiltonian $H[\varepsilon(t)]$ \cite{RZ00,ZBR98}.
The variation of $J$ with respect to $|\phi(t)\rangle$ and $|\chi(t)\rangle$ gives
Schr\"odinger's equations,
\begin{eqnarray}
  i\hbar{d\over dt}|\phi(t)\rangle&=&H[\varepsilon(t)]|\phi(t)\rangle,\\
  i\hbar{d\over dt}|\chi(t)\rangle&=&H[\varepsilon(t)]|\chi(t)\rangle.
\end{eqnarray}
Here we impose the following boundary conditions 
\begin{equation}
|\phi(0)\rangle=|\varphi_i\rangle,
\quad 
|\chi(T)\rangle=|\varphi_f\rangle.
\end{equation}
Another variation of $J$ with respect to $\varepsilon(t)$ 
gives an expression for the external field
\begin{equation}
\label{eqn:field}
  \varepsilon(t)=\frac{1}{\alpha\hbar}{\rm Im}\left[
    \langle \phi(t)| \chi(t) \rangle
    \langle \chi(t)|
     {\partial H[\varepsilon(t)]\over\partial\varepsilon(t)} | \phi(t) \rangle
   \right].
\end{equation}
In actual numerical calculations,
we usually solve these equations
with some iteration procedure \cite{ZBR98} because they are 
nonlinear with respect to $|\phi(t)\rangle$ and $|\chi(t)\rangle$.
The optimal field, Eq.~(\ref{eqn:field}), is finally given
after a local maximum of the functional is reached.

In the following subsections, we numerically demonstrate 
to control multi-level-multi-level transition problems
in quantum chaos systems:
one is a random matrix system, 
and the other is a quantum kicked rotor.

\subsection{Controlled Random Matrix System\label{sec:oct-random}}

\begin{figure}
\begin{center}
\includegraphics[scale=0.4]{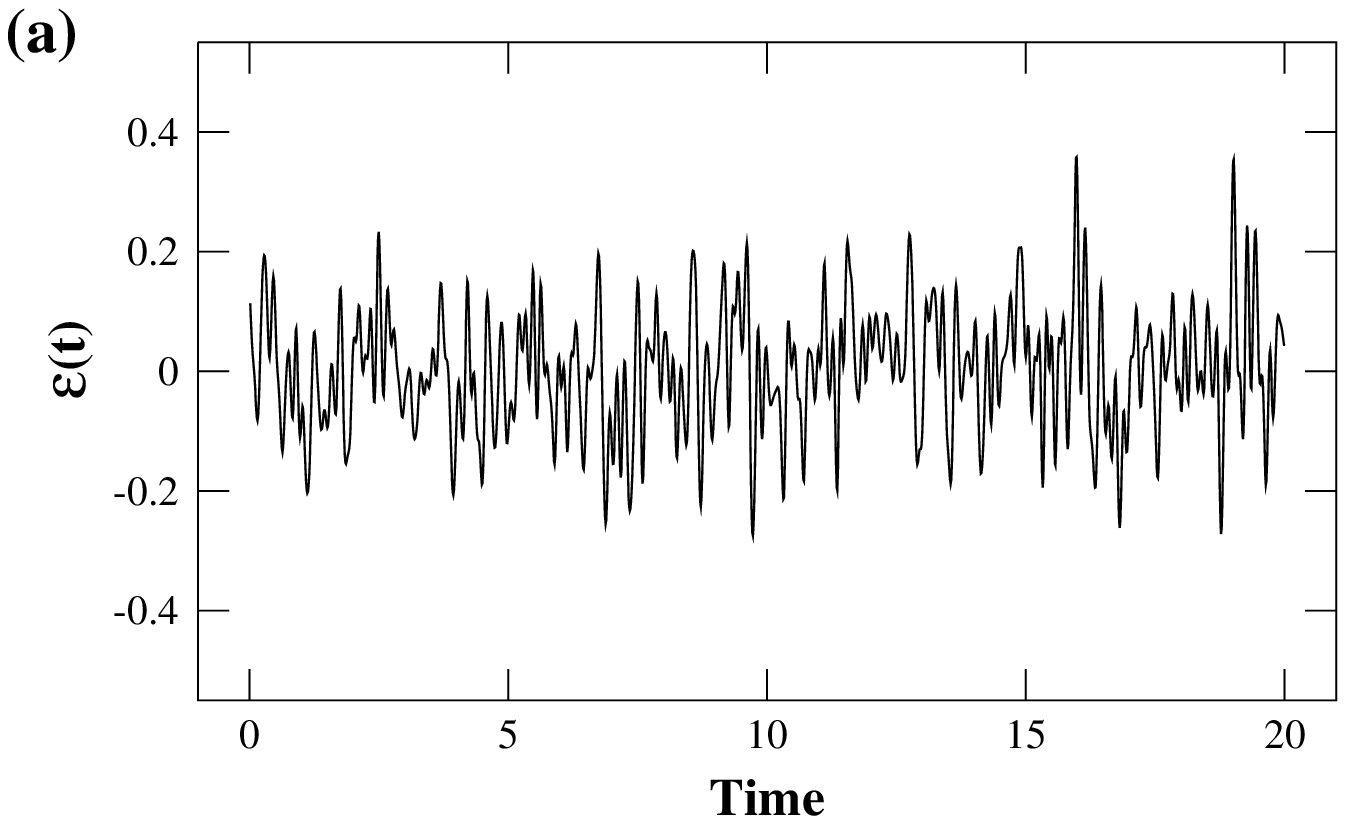}\\
\includegraphics[scale=0.4]{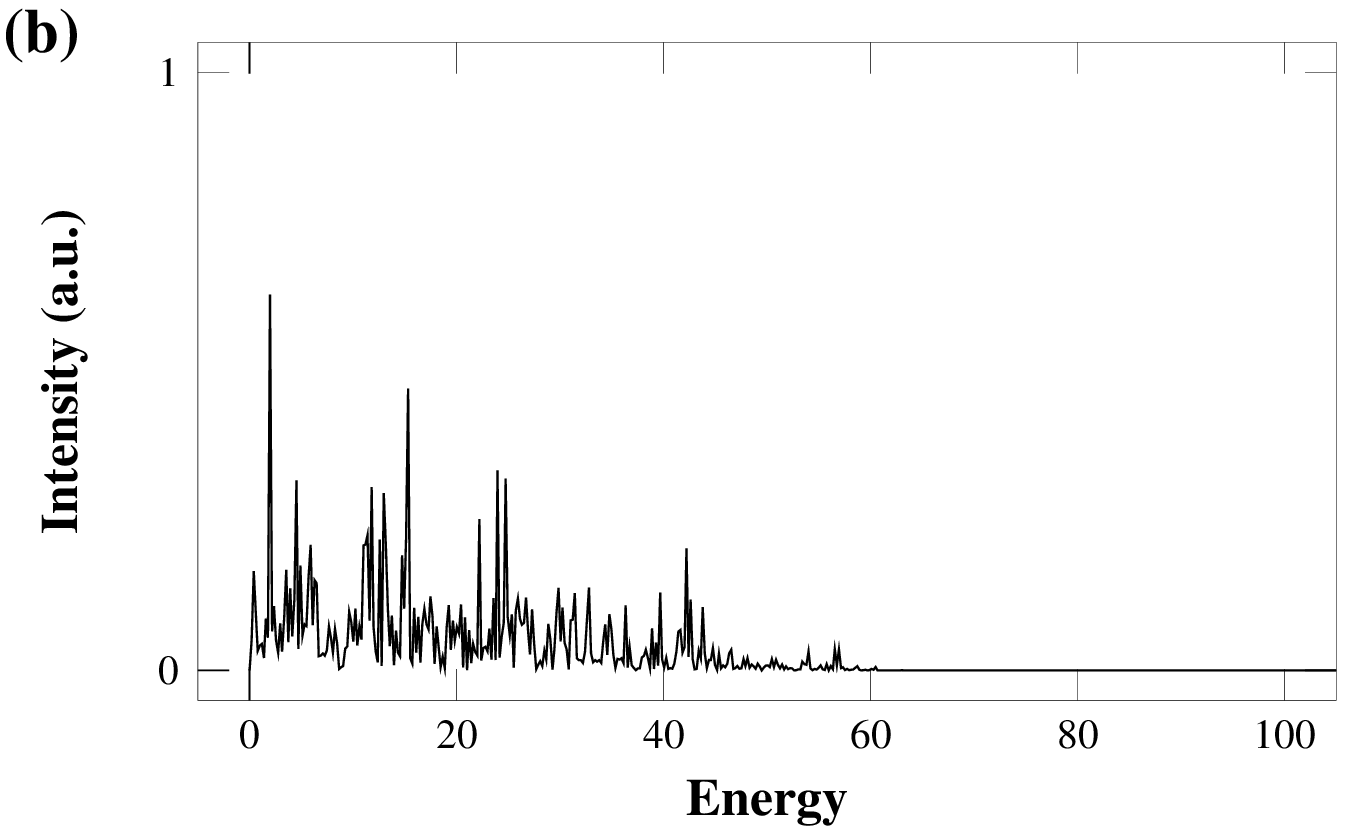}\\
\includegraphics[scale=0.4]{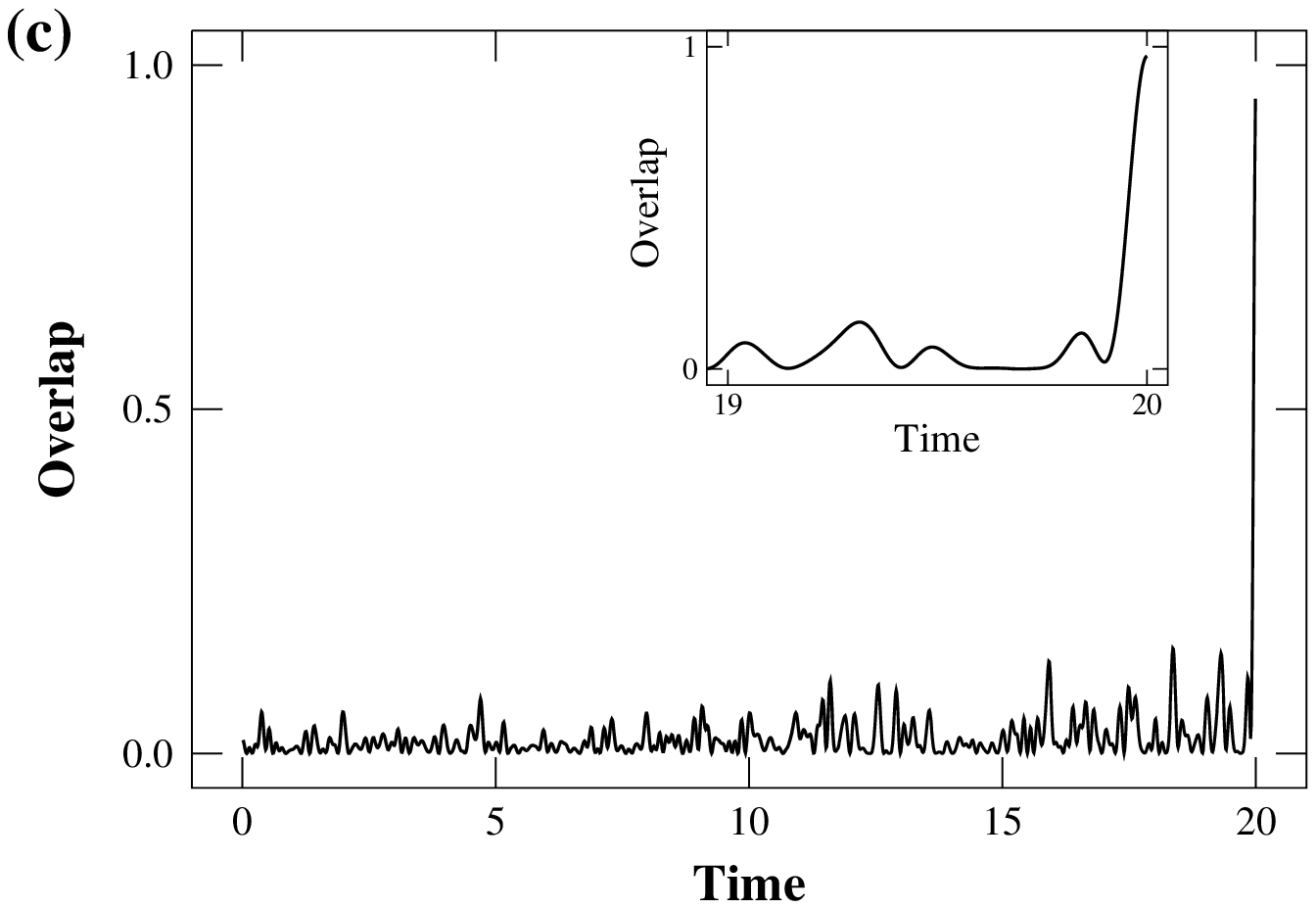}\\
\includegraphics[scale=0.4]{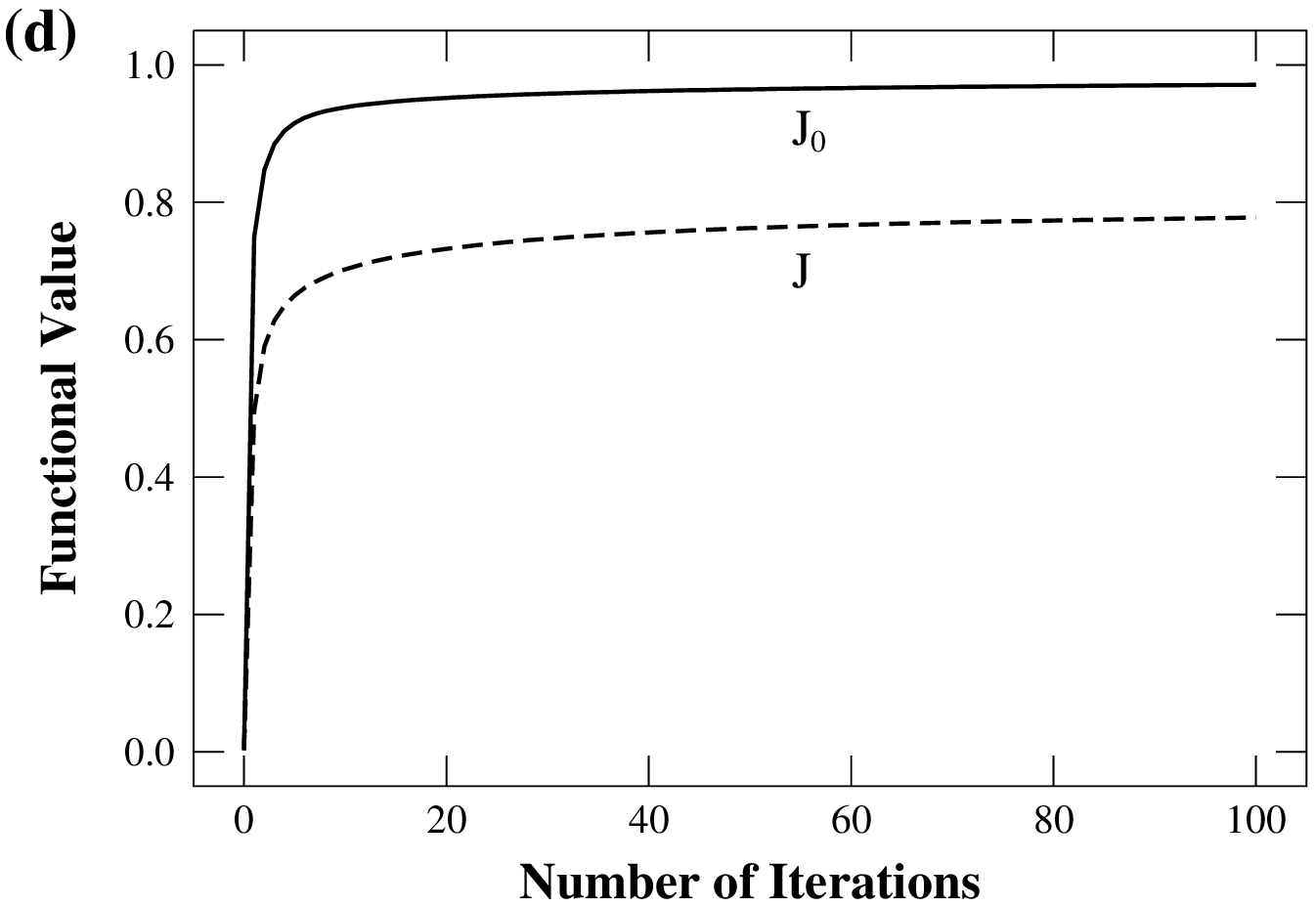}
\caption{
\label{fig:rm020}
Optimal Control between Gaussian random vectors
in a $64\times64$ random matrix system
by the Zhu-Botina-Rabitz scheme with $T=20$ and $\alpha=1$:
(a) the optimal field after 100 iterations; (b) its power spectrum;
(c) the optimal evolution of the squared overlap with the target
$\left|\langle\phi(t)|\varphi_f\rangle\right|^2$
as well as its magnified values near the target time in the inset;
(d) the convergence behavior of the overlap $J_0$ (solid)
and the functional $J$ (dashed) versus the number of iteration steps.
}
\end{center}
\end{figure}

\begin{figure}
\begin{center}
\includegraphics[scale=0.4]{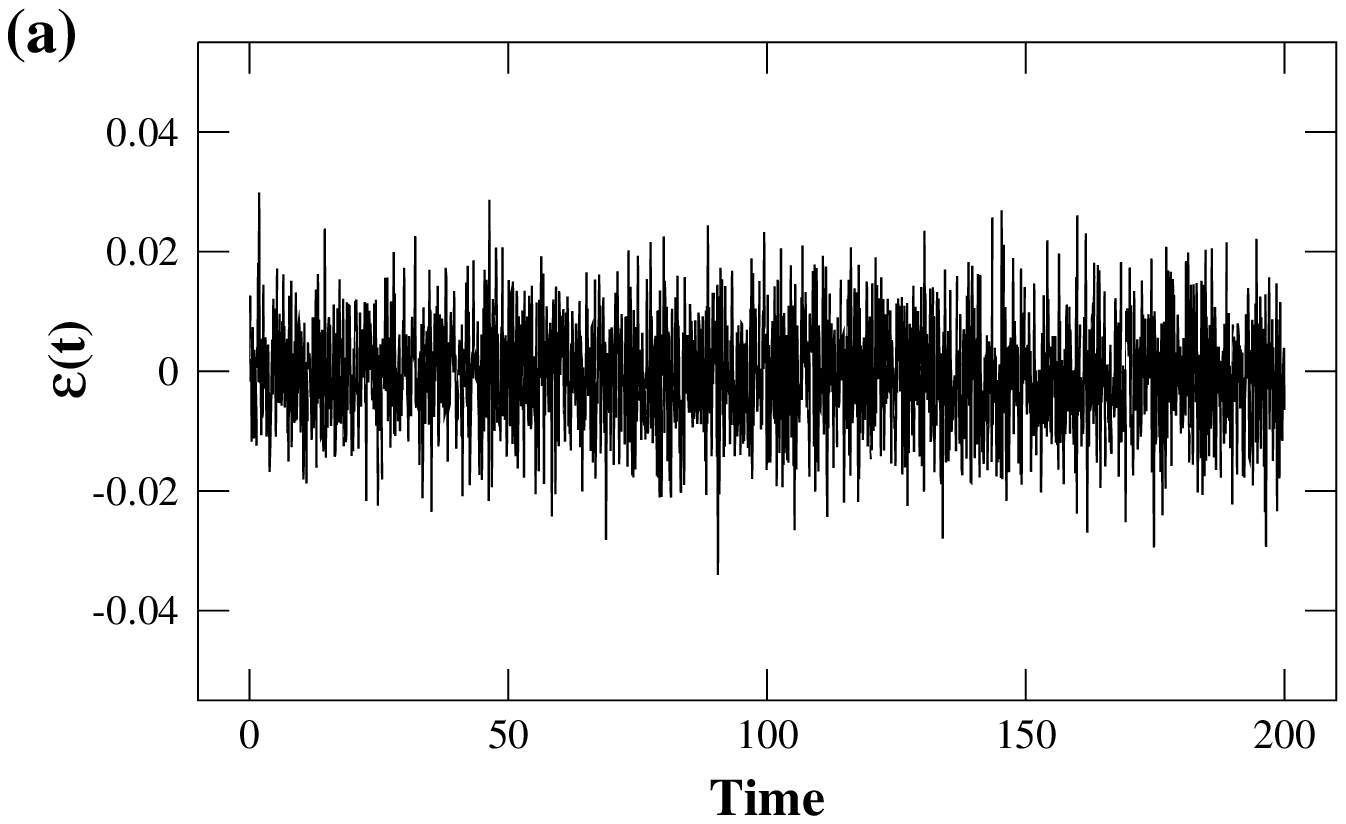}\\
\includegraphics[scale=0.4]{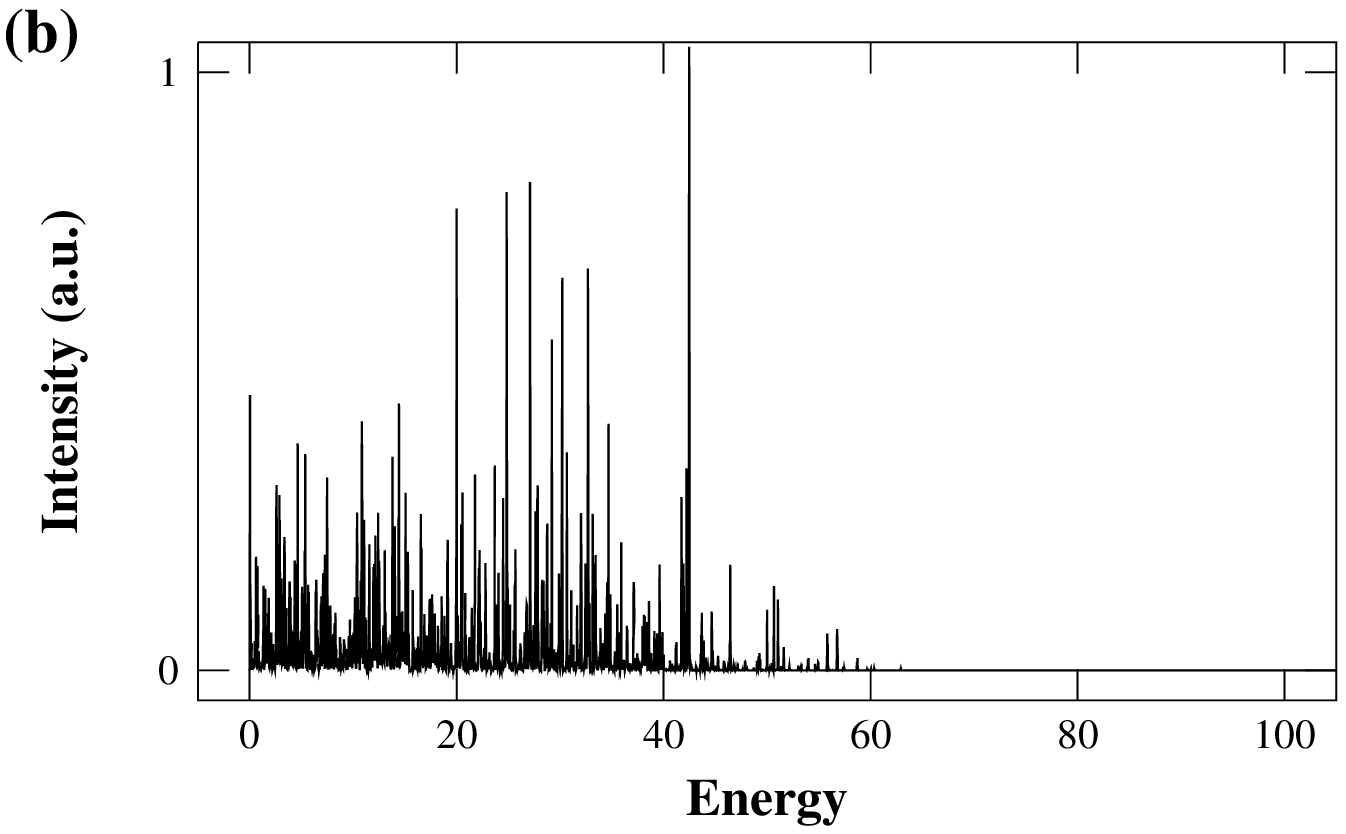}\\
\includegraphics[scale=0.4]{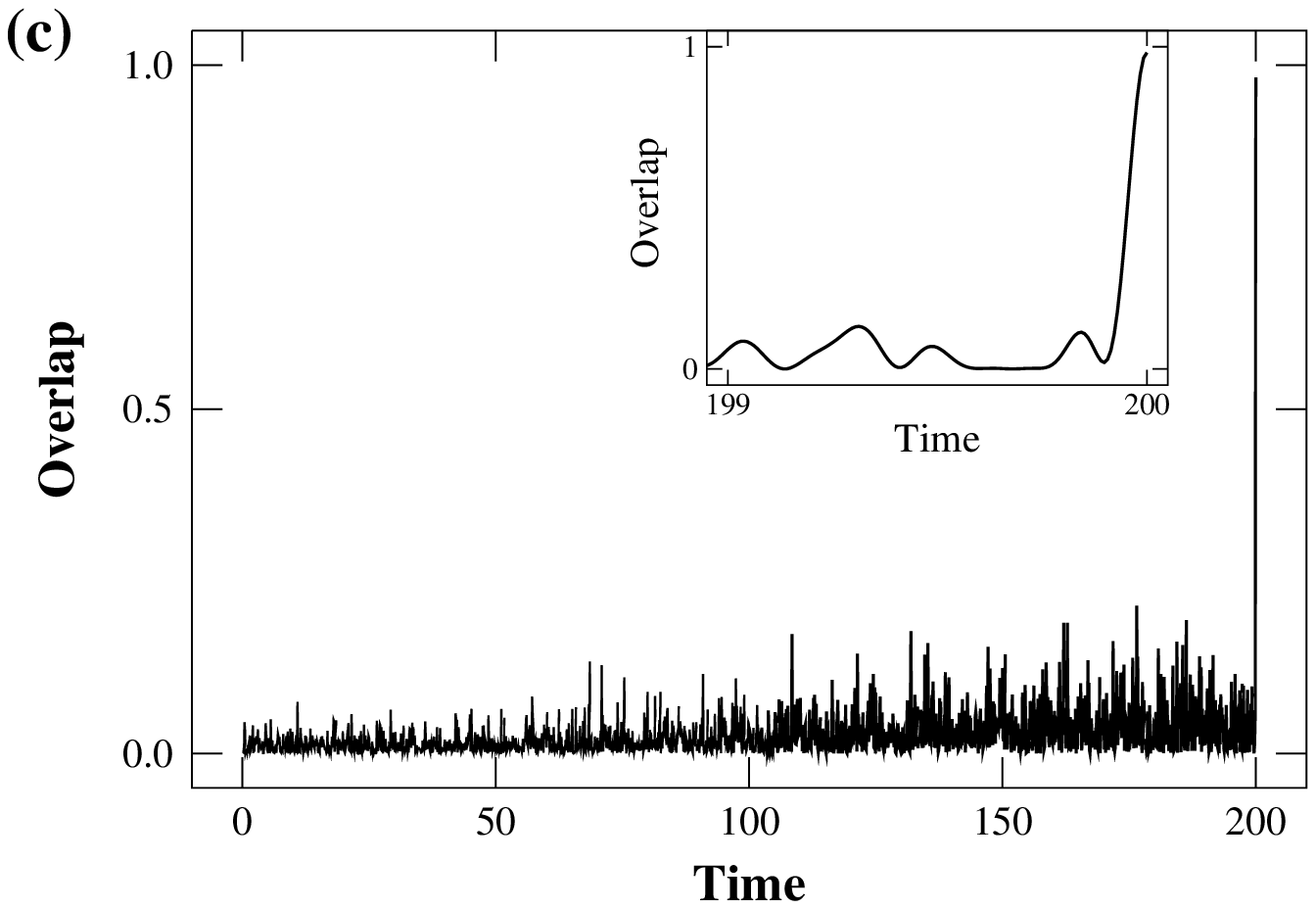}\\
\includegraphics[scale=0.4]{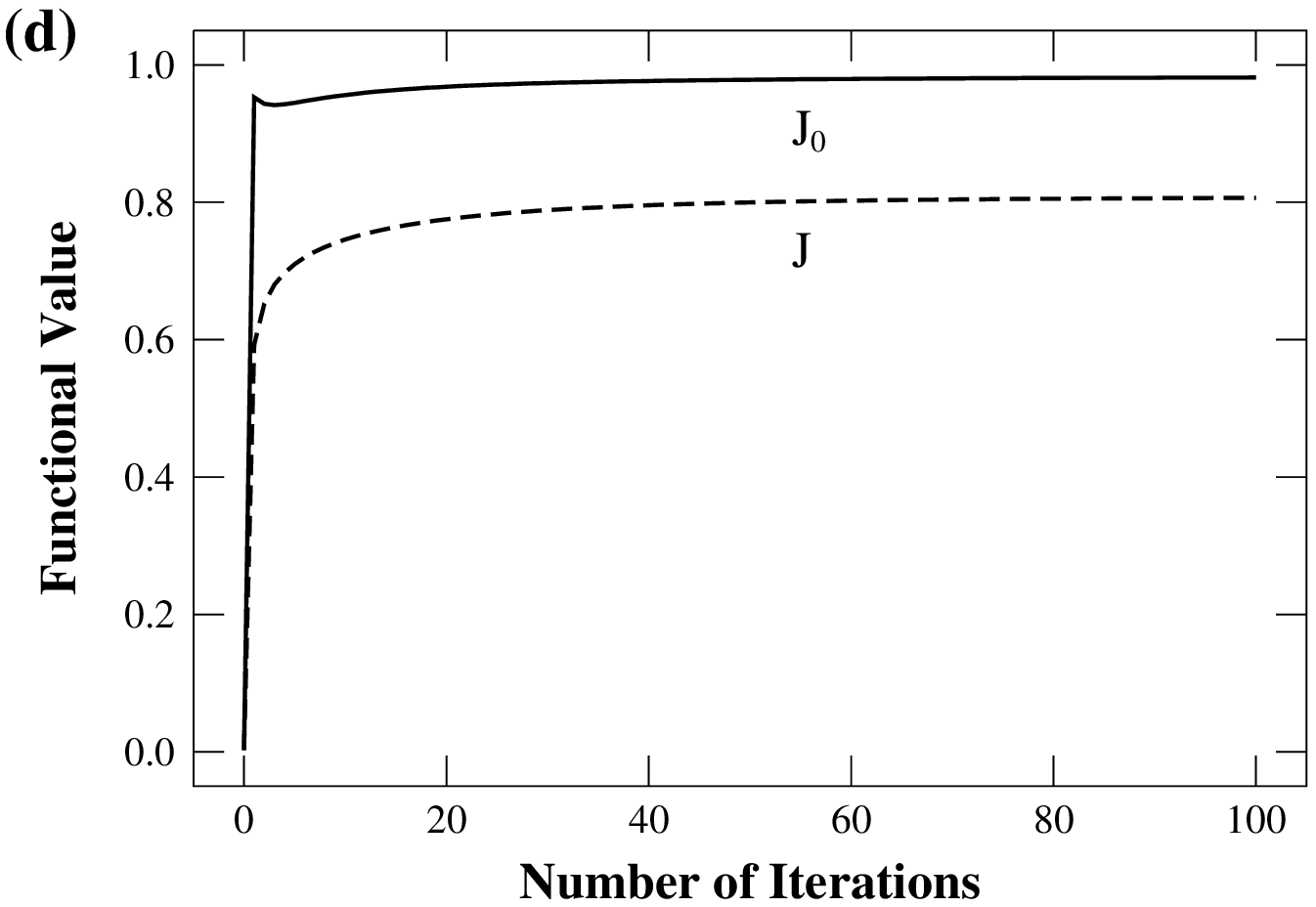}
\caption{
\label{fig:rm200}
Optimal Control between Gaussian random vectors
in a $64\times64$ random matrix system
by the Zhu-Botina-Rabitz scheme with $T=200$ and $\alpha=10$:
(a) the optimal field after 100 iterations; (b) its power spectrum;
(c) the optimal evolution of the squared overlap with the target,
$\left|\langle\phi(t)|\varphi_f\rangle\right|^2$
as well as its magnified values near the target time in the inset;
(d) the convergence behavior of the overlap $J_0$ (solid)
and the functional $J$ (dashed) versus the number of iteration steps.
}
\end{center}
\end{figure}

The random matrix was first introduced by E.P.~Wigner 
as a model to mimick unknown interactions in nuclei,
and has been studied to describe statistical natures 
of spectral fluctuations in quantum chaos systems \cite{Haake01}.
Here, we introduce a random matrix system
driven by a time-dependent external field $\varepsilon(t)$,
which is considered as a model of highly excited atoms or molecules
under an electromagnetic field.
We write the Hamiltonian
\begin{equation}
\label{eqn:random-Hamiltonian}
  H[\varepsilon(t)]=H_0+\varepsilon(t)V
\end{equation}
where $H_0$ and $V$ are $N\times N$ random matrices
subject to the Gaussian Orthogonal Ensemble (GOE),
which represent generic quantum systems with time-reversal symmetry.
The matrix elements of $H_0$ and $V$ are scaled
so that the nearest-neighbor spacing of eigenvalues of $H_0$
and the variance of the off-diagonal elements of $V$ become both unity.

Once we fix the initial state $|\varphi_i\rangle$ and
the final state $|\varphi_f\rangle$,
the optimal field $\varepsilon(t)$ is obtained by some numerical procedures
for appropriate values of the target time $T$ and the penalty factor $\alpha$.
Though there should be many situations corresponding to the choice of
$|\varphi_i\rangle$ and $|\varphi_f\rangle$,
we only consider the case where they are {\it Gaussian random vectors}.
It is defined by
\begin{equation}
  |\varphi\rangle=\sum_jc_j|\phi_j\rangle,
\end{equation}
where $c_j$ are complex numbers determined from the following 
Gaussian distribution, 
\begin{equation}
  P(c_j)\propto\exp\left(-|c_j|^2\right),
\end{equation}
and $|\phi_j\rangle$ is an orthonormal basis.\footnote{
In numerical studies below, we obtain such vectors with normalization
after generating random complex elements subject to the distribution.}
We take this state because it is typical in a random matrix system.

We show two numerical examples for a $64\times 64$ random matrix Hamiltonian:
One is the relatively short-time case with $T=20$ and $\alpha=1$
shown in Fig.~\ref{fig:rm020},
and the other is the case with $T=200$ and $\alpha=10$
shown in Fig.~\ref{fig:rm200}.
In both cases,
we obtain the optimal field $\varepsilon(t)$ after 100 iterations
using the Zhu-Botina-Rabitz (ZBR) scheme \cite{ZBR98}
with $\varepsilon(t)=0$ as an initial guess of the field.
The initial and the target state is chosen as Gaussian random vectors
as mentioned above. 
The final overlaps are $J_0=0.971$ and $0.982$, respectively.

One sees that the ZBR scheme is effective enough for random matrix systems,
i.e., the optimal fields can be obtained
even for this type of complicated problems of multi-level-multi-level 
transitions.
However, it seems that the further analysis is difficult
because the power spectra for the optimal fields,
Figs.~\ref{fig:rm020}(b) and \ref{fig:rm200}(b),
are very ``complex'', i.e., they contain many frequency components.\footnote{
In the insets of Figs.~\ref{fig:rm020}(c) and \ref{fig:rm200}(c),
we show the overlaps $|\langle\phi(t)|\varphi_f\rangle|^2$ near $t=T$
in a magnified scale.
They exhibit almost the same curves
in spite of the different optimal fields.
This is because the optimal field is small enough
so that the dynamics is not affected in this time scale.}

\subsection{Controlled Quantum Kicked Rotor\label{sec:kr}}

\begin{figure}
\begin{center}
\includegraphics[scale=0.4]{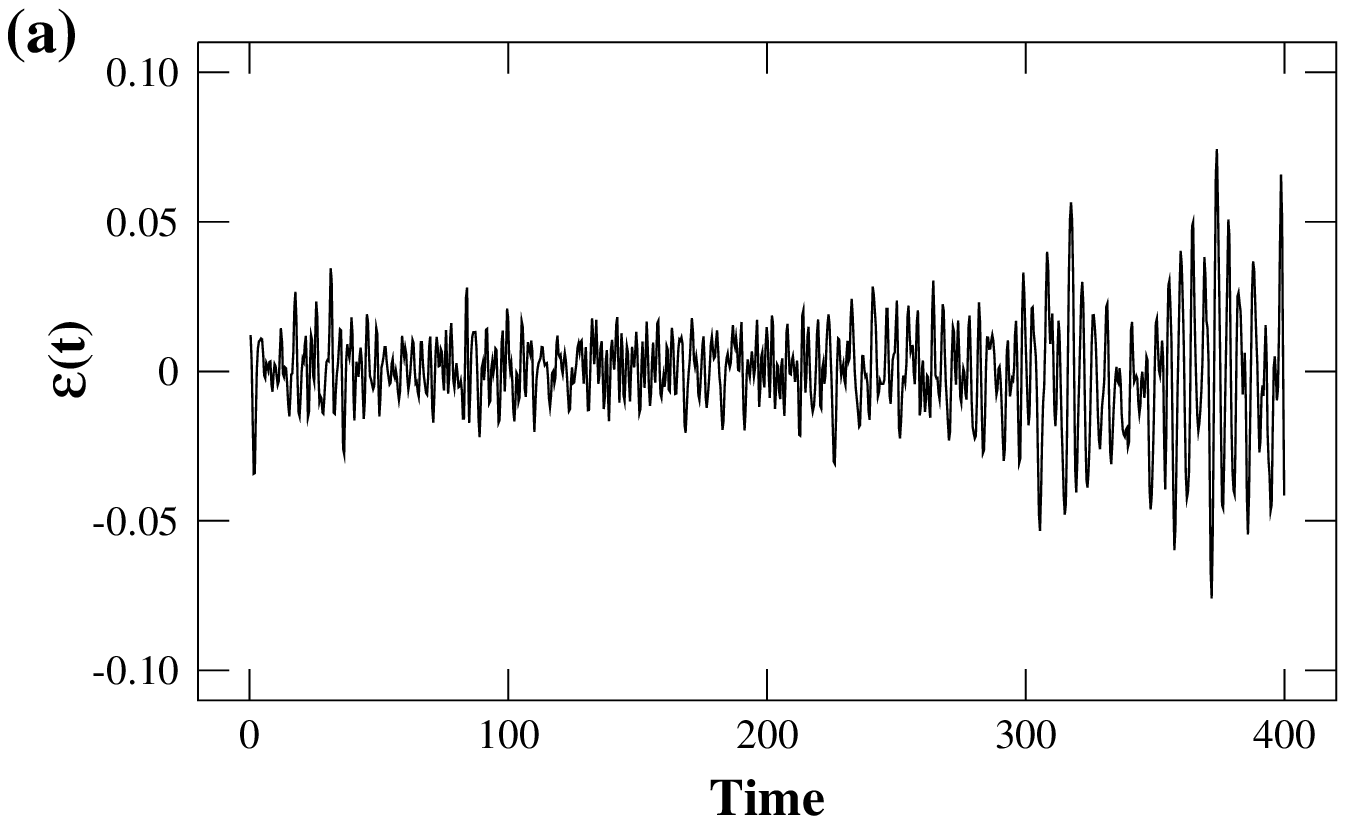}\\
\includegraphics[scale=0.4]{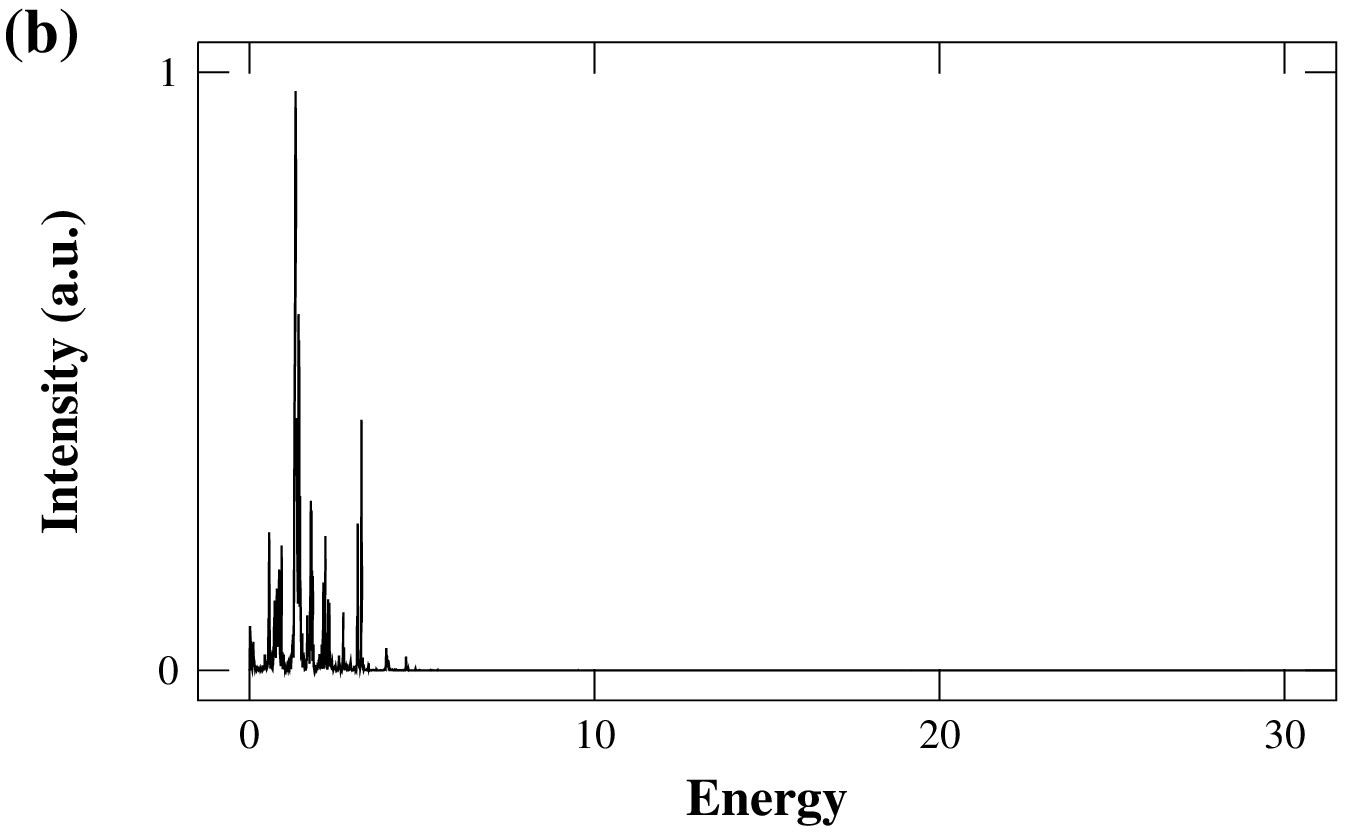}\\
\includegraphics[scale=0.4]{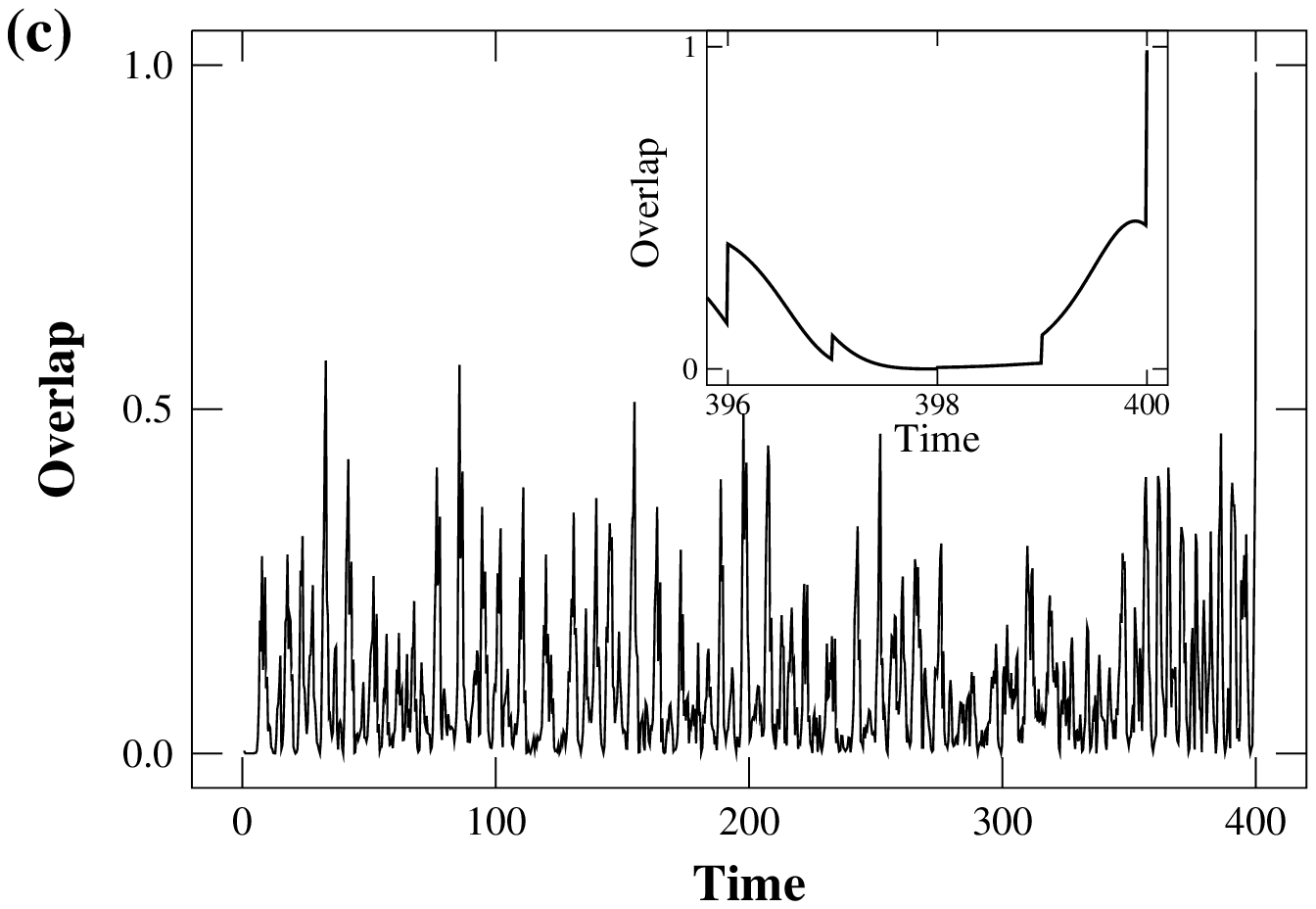}\\
\includegraphics[scale=0.4]{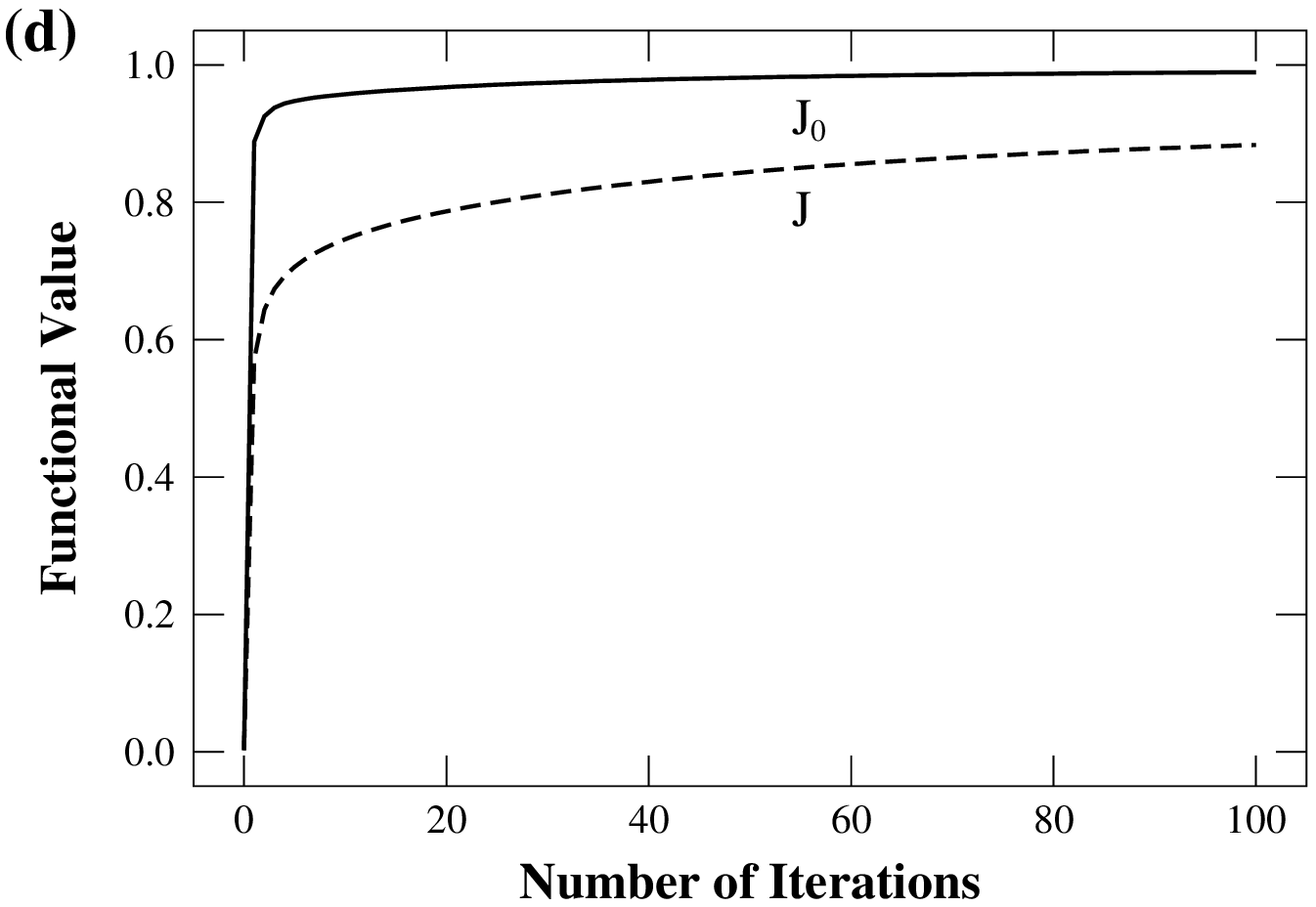}
\caption{
\label{fig:fdK1}
Optimal Control in a regular kicked rotor with $K=1$ and $\hbar=0.3436$
by the Zhu-Botina-Rabitz scheme with $T=400$ and $\alpha=1$:
(a) the optimal field after 100 iterations; (b) its power spectrum;
(c) the optimal evolution of the squared overlap with the target
$\left|\langle\phi(t)|\varphi_f\rangle\right|^2$
as well as its magnified values near the target time in the inset;
(d) the convergence behavior of the overlap $J_0$ (solid)
and the functional $J$ (dashed) versus the number of iteration steps.
}
\end{center}
\end{figure}
\begin{figure}
\begin{center}
\includegraphics[scale=0.4]{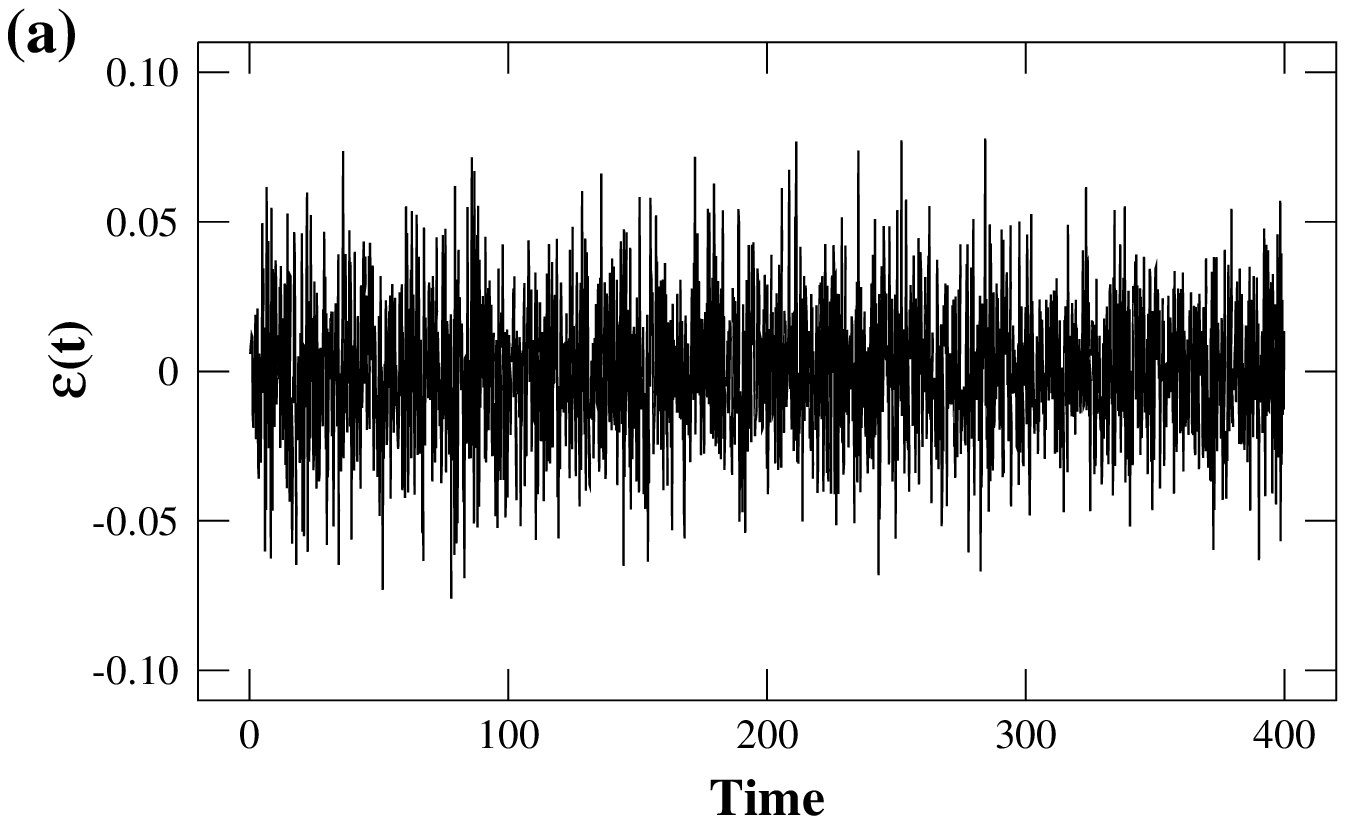}\\
\includegraphics[scale=0.4]{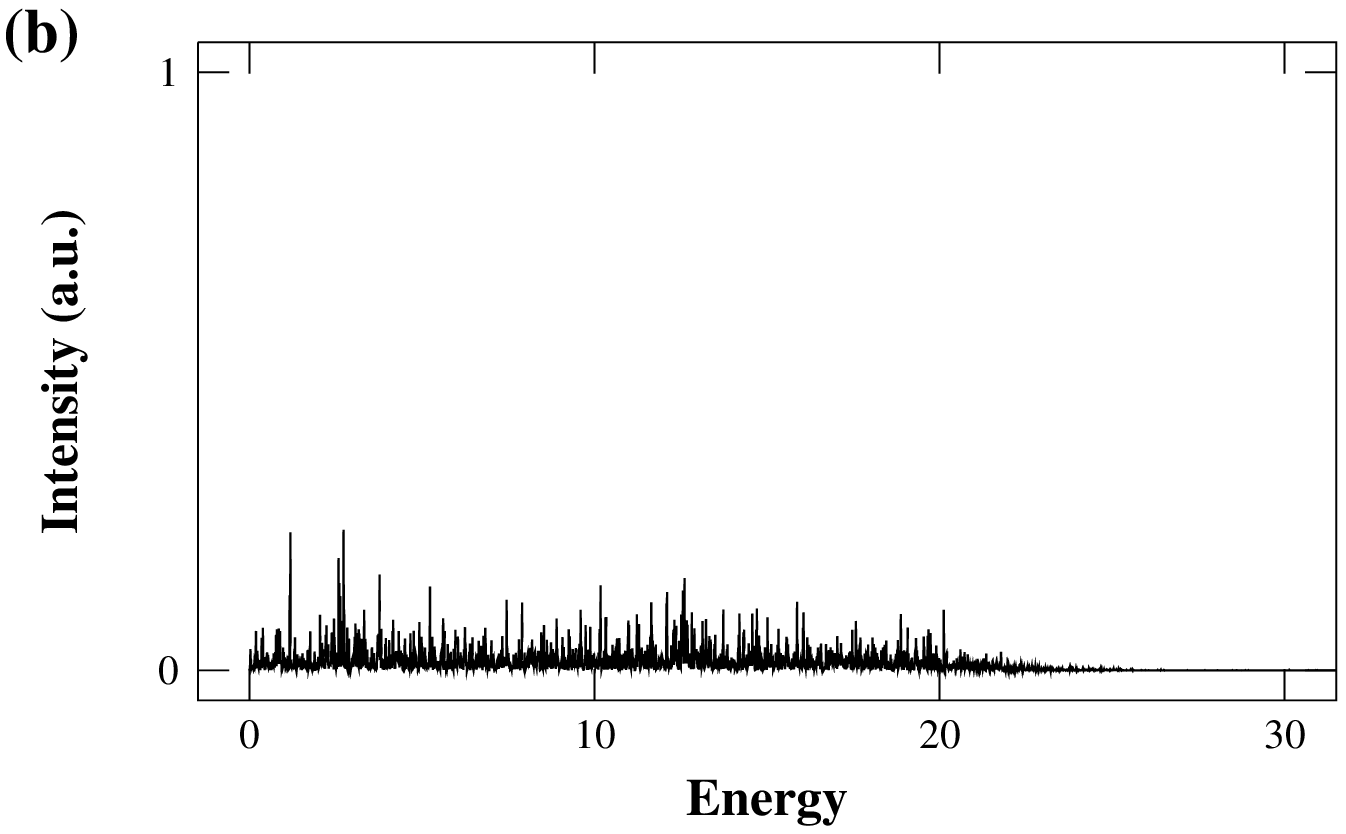}\\
\includegraphics[scale=0.4]{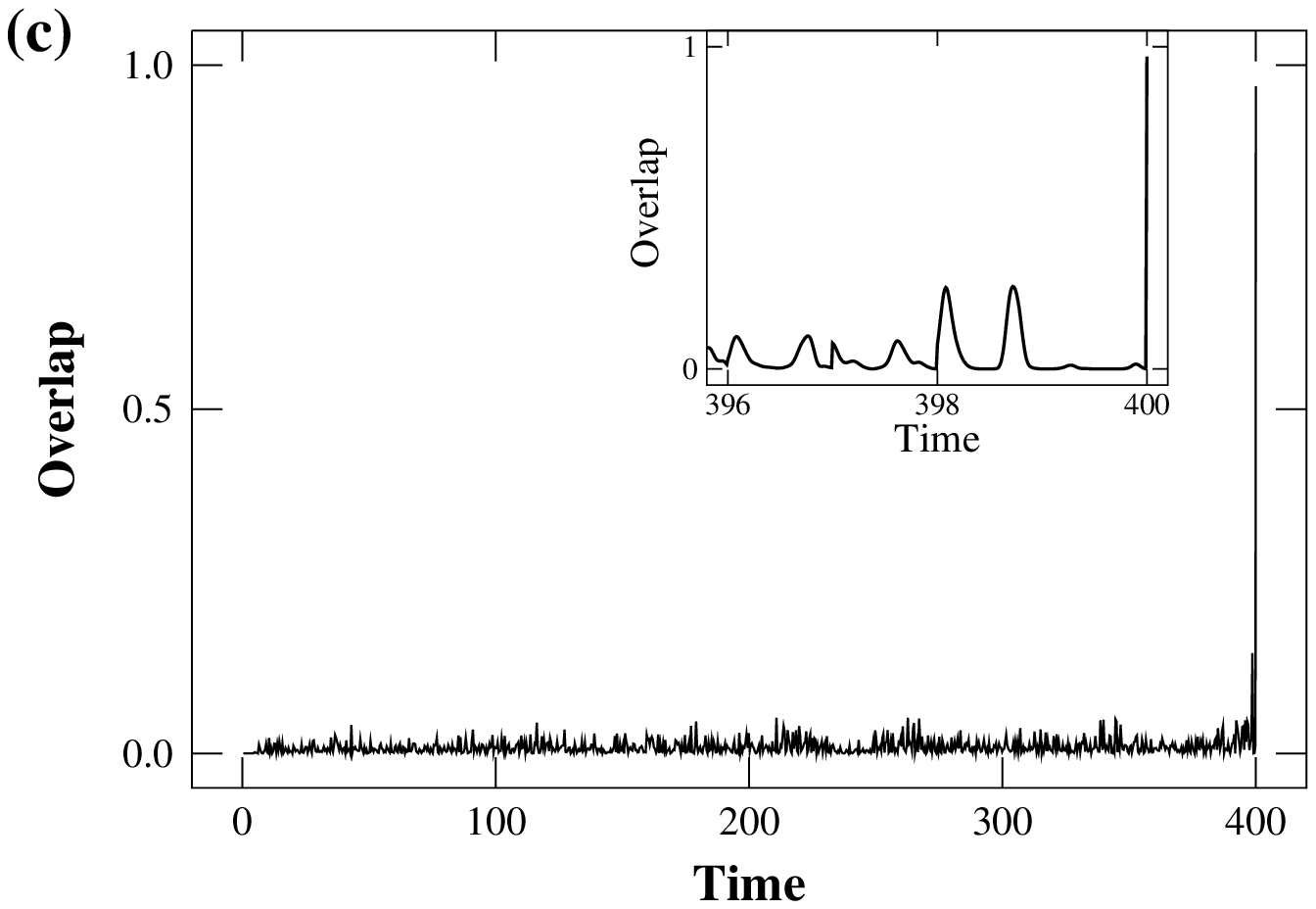}\\
\includegraphics[scale=0.4]{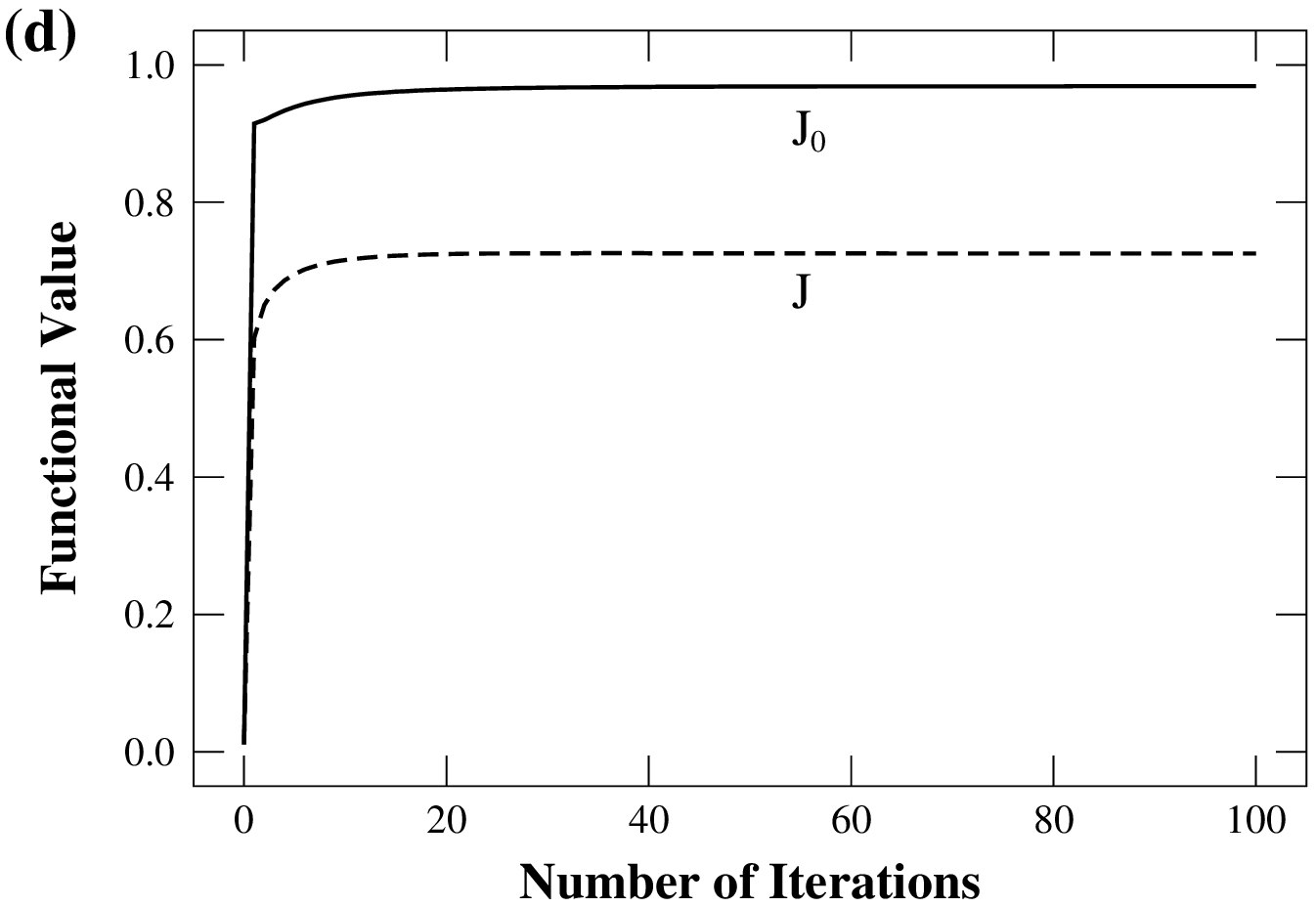}
\caption{
\label{fig:fdK7}
Optimal Control in a chaotic kicked rotor with $K=7$ and $\hbar=0.3436$
by the Zhu-Botina-Rabitz scheme with $T=400$ and $\alpha=1$:
(a) the optimal field after 100 iterations; (b) its power spectrum;
(c) the optimal evolution of the squared overlap with the target,
$\left|\langle\phi(t)|\varphi_f\rangle\right|^2$
as well as its magnified values near the target time in the inset;
(d) the convergence behavior of the overlap $J_0$ (solid)
and the functional $J$ (dashed) versus the number of iteration steps.
}
\end{center}
\end{figure}

The kicked rotor (or the standard map)
is one of famous models in chaotic dynamical systems,
and has been studied in various situations \cite{Haake01}.
One feature of its chaotic dynamics is the {\it deterministic diffusion} 
along the momentum direction.
It is also well known that, 
if we quantize this system, this diffusion is suppressed by 
the effects of the wavefunction localization in momentum space \cite{Gutzwiller90}.

Here we employ the quantum kicked rotor 
as a simple model of quantum chaos systems.
The Hamiltonian of a kicked rotor is written as
\begin{equation}
  H_{\rm KR}(t)=\frac{p^2}{2}
      +\frac{K}{\tau}\cos\theta
         \sum_{n=-\infty}^{\infty}\delta(t-n\tau),
\end{equation}
where $\theta$ is an angle (mod $2\pi$), 
$p$ momentum, 
$K$ a kick strength, and $\tau$ a period between kicks.
An external field $\varepsilon(t)$ is applied through the coupling Hamiltonian
\begin{equation}
  H_{\rm I}[\varepsilon(t)]=-\mu(\theta)\varepsilon(t),
\end{equation}
where the dipole moment is assumed
to be 
\begin{equation}
\mu(\theta)=-\cos(\theta+\delta\theta_0).
\end{equation}
The extra phase $\delta\theta_0$ is introduced to
break symmetry of the system.
We take $\delta\theta_0=\pi/3$ in the numerical calculations
throughout this paper.
The total Hamiltonian
is given by
\begin{equation}
H[\varepsilon(t)]=H_{\rm KR}(t)+H_{\rm I}[\varepsilon(t)].
\end{equation}

For easiness of computation, we impose a periodic boundary condition
for $p$ as well as $\theta$; the phase space of the corresponding
classical system becomes a two-dimensional torus~\cite{Iz86,CS86}.
In this case, Planck's constant is given by $\hbar=2\pi M/\tau N$,
where $p=\pm M\pi$ defines the periodic boundaries in the momentum space,
and $N$ is the number of discrete points describing $\theta$ and $p$.
In the actual calculations, we set $\tau=1$.

The kicked rotor is often described only at
discrete time immediately after/before the periodic kicks.
In our control problem, however,
we must represent dynamics driven by $\varepsilon(t)$ between those kicks.
Then, we can apply Zhu-Botina-Rabitz scheme as usual.
According to Eq.~(\ref{eqn:field}),
the optimal external field is given by
\begin{equation}
\label{eqn:kr-field}
   \varepsilon(t)=-\frac{1}{\alpha\hbar}
     {\rm Im}\left[
        \langle \phi(t)| \chi(t) \rangle
        \langle \chi(t)| \mu(\theta) | \phi(t) \rangle
     \right].
\end{equation}
Note that, because $\mu(\theta)$ commutes
with the unitary operator $e^{-i K\cos\theta/\hbar}$ of a kick,
$\varepsilon(t)$ is obtained as a continuous function of time
even at the moment of the delta kicks.

In Figs.~\ref{fig:fdK1} and \ref{fig:fdK7}, 
we show numerical results for the quantum kicked rotor\footnote{
We use the parameters $N=128$ and $M=7$. 
Thus the quantum states are represented by
$128$ discrete points, and the range of momentum is from $-7\pi$ to $7\pi$.
The value of $\hbar=2 \pi M/N$ is $0.3436$.}
as in Sec.~\ref{sec:oct-random}.
The system parameters are chosen to pick up a regular dynamics (Fig.~\ref{fig:fdK1})  
and a chaotic dynamics (Fig.~\ref{fig:fdK7}), and the others 
are $T=400$ and $\alpha=1$.
The optimal field after 100 iterations 
for the regular case, Fig.~\ref{fig:fdK1}(a), is 
rather simpler than that for the chaotic case, Fig.~\ref{fig:fdK7}(a).
[See also Fig.~\ref{fig:fdK1}(b) and \ref{fig:fdK7}(b).] 
This is because more states are involved in the latter chaotic process.

\begin{figure}
\begin{center}
\includegraphics[scale=0.25]{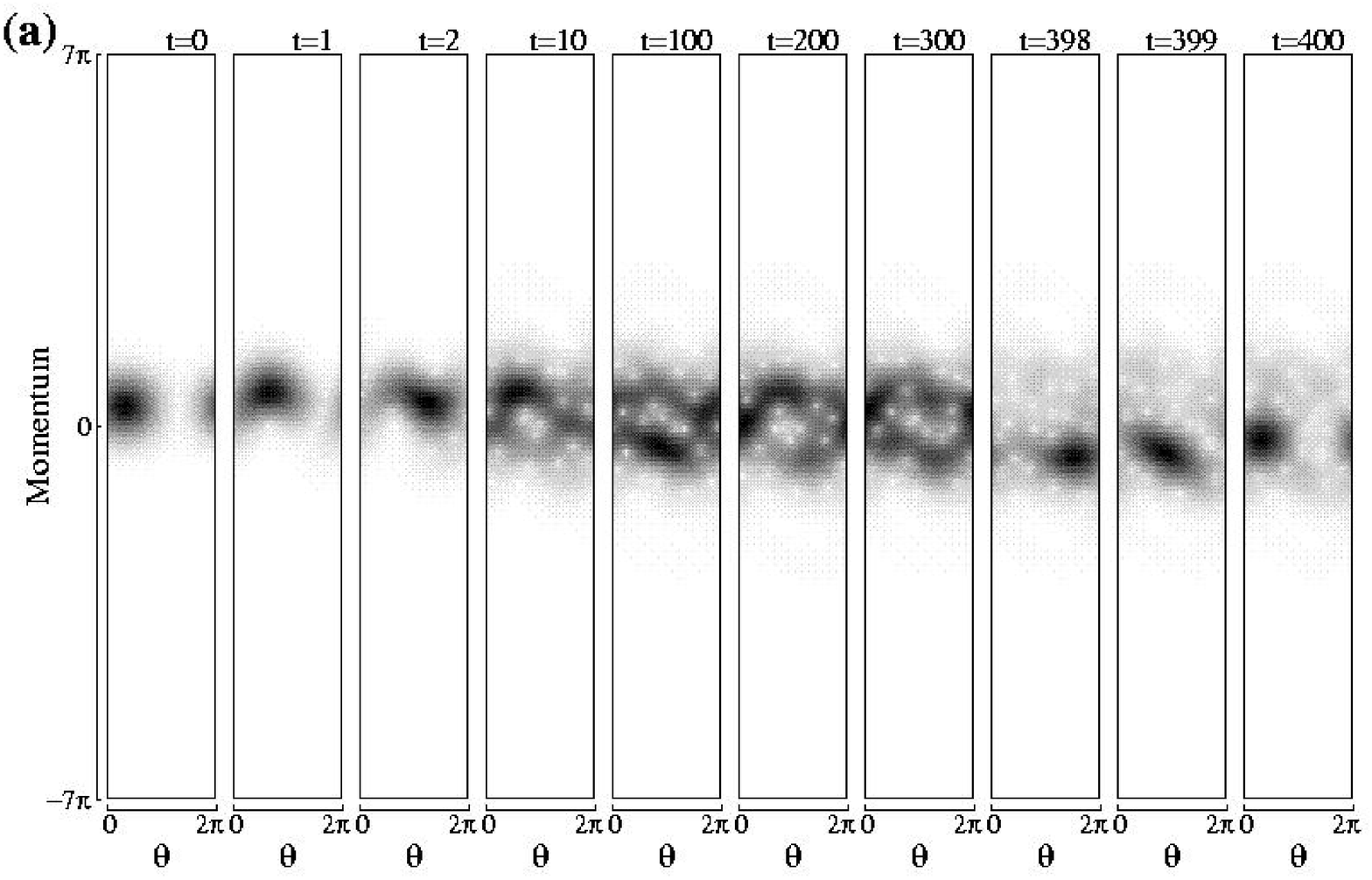}
\includegraphics[scale=0.25]{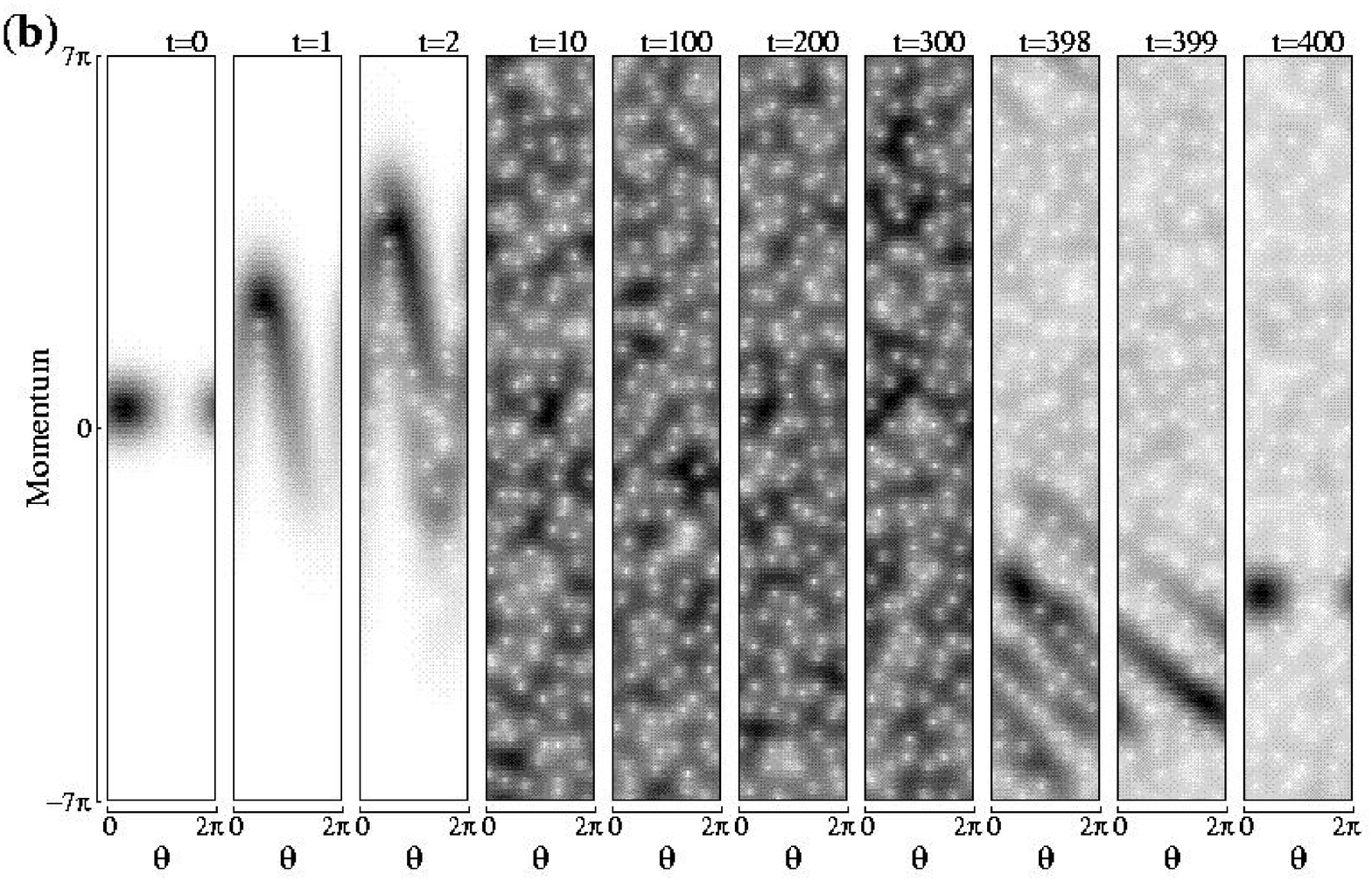}
\caption{
\label{fig:krwp}
Time evolution of the Husimi distribution for quantum kicked rotors
with $\hbar=0.3436$ under an optimal field after 100 iterations.
The Zhu-Botina-Rabitz scheme was used with
the penalty factor $\alpha=1$ and the target time $T=400$.
From left to right, quantum states
immediately after the kick at $t=0$, $1$, $2$, $10$, $100$, $200$, $300$,
$398$, $399$, and $400$ are depicted.
(a) The parameters are $K=1$ (regular case), 
$(\theta_i,p_i)=(1.0,1.0)$ and $(\theta_f,p_f)=(1.0,-1.0)$;
(b) $K=7$ (chaotic case), 
$(\theta_i,p_i)=(1.0,1.0)$ and $(\theta_f,p_f)=(1.0,-10.0)$.
}
\end{center}
\end{figure}

Next we investigate the wavepacket dynamics in phase space
using the Husimi representation \cite{Takahashi89}.
The initial and final states,
$|\varphi_i\rangle$ and $|\varphi_f\rangle$,
are chosen as minimum uncertainty (Gaussian) packets
centered at $(\theta_i,p_i)$ and $(\theta_f,p_f)$, respectively. %, i.e.,
In Fig.~\ref{fig:krwp}(a), we show the result for the regular 
case corresponding to Fig.~\ref{fig:fdK1}. 
Optimal control is achieved for a wavepacket motion within a torus
with $J_0=0.989$.\footnote{
When the control purpose is to steer a wavepacket in 
a torus to another place in another torus, 
OCT fails. This is because the wavepacket is trapped in one torus and 
it is very hard to escape from the torus with a weak external field.
}
Figure \ref{fig:krwp}(b) shows the controlled dynamics 
for the chaotic case corresponding to Fig.~\ref{fig:fdK7}. 
In this case, the wavepacket once spreads all over the phase space due 
to the chaotic nature of the system,
but it gets together at the target time $T$ with $J_0=0.969$.
In both regular and chaotic cases, 
the ZBR-OCT scheme works well for the quantum kicked rotor \cite{TFM04}.

\section{Coarse-Grained Picture\label{sec:cg-picture}}

If we apply a resonant external field to a two-level system,
we can observe a Rabi oscillation.
In such a case, the quantum state is well described by 
\begin{eqnarray}
\label{eqn:usual-Rabi-state}
  |\phi(t)\rangle
    &=&e^{E_1 t/i\hbar}|\phi_1\rangle\cos|\Omega|t\nonumber\\
    &&\qquad -ie^{-i\theta}e^{E_2 t/i\hbar}|\phi_2\rangle\sin|\Omega|t,
\end{eqnarray}
where $|\phi_1\rangle$ and $|\phi_2\rangle$ 
($E_1$ and $E_2$) are two eigenstates (eigen-energies) of the system,
$|\Omega|\equiv|\varepsilon_0\mu_{12}|/\hbar$ the Rabi frequency,
$\mu_{12}\equiv\langle\phi_1|\hat\mu|\phi_2\rangle$ 
matrix elements of a dipole operator $\hat\mu$,
$\varepsilon_0$ an amplitude of the field,
and $\theta$ a certain phase parameter.

In this section, we study the controlled dynamics
from an initial state $|\varphi_i\rangle$ at $t=0$ to
a target state $|\varphi_f\rangle$ at $t=T$ in
a multi-state quantum mechanical system described by Eq.~(\ref{eqn:random-Hamiltonian}).
By introducing a ``coarse-grained'' picture,
which means neglecting highly oscillating terms as the case 
of rotating-wave approximation (RWA) \cite{AE87} and 
assuming that $|\varphi_i\rangle$ and $|\varphi_f\rangle$ contain
many eigenstates without any correlation between them,
we show that the controlled dynamics can be represented
as a transition between a pair of time-dependent states \cite{TF04}. 

\subsection{Coarse-Grained Rabi State and Frequency\label{sec:CG-Rabi}}

As shown in Sec.~\ref{sec:oct-random}, 
the overlap in the controlled dynamics 
rapidly oscillates because the system contains many states.
To analyze this complicated behavior more easily, 
we introduce the following two time-dependent states,
\begin{equation}
\label{eqn:basis}
  |\phi_0(t)\rangle=\hat U_0(t,0)|\varphi_i\rangle,\quad
  |\chi_0(t)\rangle=\hat U_0(t,T)|\varphi_f\rangle
\end{equation}
where 
\begin{equation}
\hat U_0(t_2,t_1)=e^{-iH_0(t_2-t_1)/\hbar}
\end{equation}
is a ``free''-propagator with $H_0$
from $t=t_1$ to $t_2$, and $T$ is a target time.
These states are an analogue of eigenstates
in the usual Rabi state (\ref{eqn:usual-Rabi-state}),
and we try to describe the controlled dynamics
as a transition from $|\phi_0(t)\rangle$ to $|\chi_0(t)\rangle$.

We introduce another quantum state
by a linear combination of the two time-dependent states,
\begin{equation}
  |\phi(t)\rangle=|\phi_0(t)\rangle c(t)+|\chi_0(t)\rangle s(t)
\end{equation}
where $c(t)$ and $s(t)$ are functions satisfying a normalization 
condition: 
\begin{equation}
|c(t)|^2+|s(t)|^2=1.
\end{equation}
\begin{widetext}
If we require $|\phi(t)\rangle$ to satisfy Schr\"odinger's equation,
we obtain
\begin{equation}
  i\hbar\left[|\phi_0(t)\rangle{d\over dt}c(t)
     +|\chi_0(t)\rangle{d\over dt}s(t)\right]
  =\varepsilon(t)V\left[|\phi_0(t)\rangle c(t)+|\chi_0(t)\rangle s(t)\right].
\end{equation}

Multiplying $\left<\phi_0(t)\right|$ and $\left<\chi_0(t)\right|$ from the left
gives the following equations for $c(t)$ and $s(t)$
\begin{equation}
  i\hbar{d\over dt}\pmatrix{c(t)\cr s(t)\cr}
  =\pmatrix{\langle\phi_0(t)|\varepsilon(t)V|\phi_0(t)\rangle&
            \langle\phi_0(t)|\varepsilon(t)V|\chi_0(t)\rangle\cr
            \langle\chi_0(t)|\varepsilon(t)V|\phi_0(t)\rangle&
            \langle\chi_0(t)|\varepsilon(t)V|\chi_0(t)\rangle\cr}
   \pmatrix{c(t)\cr s(t)\cr}.
\label{eqn:de-cs}
\end{equation}
where we have used
\begin{equation}
|\langle\phi_0(t)|\chi_0(t)\rangle|\ll1
\end{equation}
which is satisfied when $|\varphi_i\rangle$ and $|\varphi_f\rangle$ are
random vectors with a large number of elements.

Our aim is not to solve Eq.~(\ref{eqn:de-cs}) exactly,
but to find a coarse-grained (CG) solution
by ignoring rapidly oscillating terms
when the target time $T$ is large enough.
If we use the well-optimized field $\varepsilon(t)$,
we expect that the following condition
\begin{equation}
\label{eqn:cond}
  \left|\langle\phi_0(t)|\varepsilon(t)V|\phi_0(t)\rangle\right|,\
  \left|\langle\chi_0(t)|\varepsilon(t)V|\chi_0(t)\rangle\right|
   \ll\left|\langle\phi_0(t)|\varepsilon(t)V|\chi_0(t)\rangle\right|,
\end{equation}
are satisfied for $T\rightarrow\infty$ under
the coarse-grained picture.
The validity of this condition will be 
checked in Sec.~\ref{sec:validity}.
\end{widetext}

Under this condition, we obtain the following simple equations
\begin{equation}
  i\hbar{d\over dt}\pmatrix{c(t)\cr s(t)\cr}
  =\pmatrix{0&\hbar\Omega\cr\hbar\Omega^*&0\cr}\pmatrix{c(t)\cr s(t)\cr},
\end{equation}
where
\begin{equation}
\label{eqn:omega}
  \Omega\equiv\left<
    {\langle\phi_0(t)|\varepsilon(t)V|\chi_0(t)\rangle\over\hbar}
  \right>_{\rm CG}
\end{equation}
is a frequency defined by ignoring rapidly oscillating terms.
We also expect that $\Omega$ has a constant (time-independent) value,
which will be justified below.
Then, the boundary conditions $c(0)=1$ and $s(0)=0$ gives a solution
\begin{equation}
  c(t)=\cos|\Omega|t,\qquad s(t)=-ie^{-i\theta}\sin|\Omega|t
\end{equation}
where $e^{i\theta}=\Omega/|\Omega|$.
The final expression of the controlled dynamics is
\begin{equation}
\label{eqn:cg-rabi}
  |\phi(t)\rangle=|\phi_0(t)\rangle\cos|\Omega|t
   -ie^{-i\theta}|\chi_0(t)\rangle\sin|\Omega|t.
\end{equation}
Note that this state is interpreted to represent
a transition between $|\phi_0(t)\rangle$ and $|\chi_0(t)\rangle$
or that between $|\varphi_i \rangle$ and $|\varphi_f \rangle$.
Since this is very similar to 
the usual Rabi state, Eq.~(\ref{eqn:usual-Rabi-state}),
we call this state, Eq.~(\ref{eqn:cg-rabi}), ``CG Rabi state'',
and the frequency, Eq.~(\ref{eqn:omega}), ``CG Rabi frequency''.

\subsection{Actual Coarse-Graining Procedure\label{sec:validity}}

In the previous subsection,
we have introduced the concept ``coarse-graining'' (CG) to define
the CG Rabi frequency $\Omega$, Eq.~(\ref{eqn:omega}).
In the actual calculations,
we carry out this procedure by averaging over a certain time interval,
\begin{equation}
  \left\langle A(t)\right\rangle_{\rm CG}
  \equiv{1\over t_2-t_1}\int_{t_1}^{t_2}A(t')dt'.
\end{equation}
Though this result depends on the choice of $t_1, t_2$ in general,
we consider that there exists a natural time scale 
where the time averaging is meaningful.
In optimal control problems, if we choose the target time $T$ large enough,
we can substitute the range of the integration into above expression,
i.e., $t_1=0$ to $t_2=T$.

To check when the condition, Eq.~(\ref{eqn:cond}),
is fulfilled,
and when the CG Rabi frequency $\Omega$ defined in Eq.~(\ref{eqn:omega}) becomes constant,
we introduce the following integrals
\begin{eqnarray}
  F(t)&=&
     \int_0^t\langle\phi_0(t')|\varepsilon(t')V|\chi_0(t')\rangle dt'\\
  g_\phi(t)&=&
     \int_0^t\langle\phi_0(t')|\varepsilon(t')V|\phi_0(t')\rangle dt'\\
  g_\chi(t)&=&
     \int_0^t\langle\chi_0(t')|\varepsilon(t')V|\chi_0(t')\rangle dt'.
\end{eqnarray}
Though the integrands are rapidly oscillating,
a certain smoothness can be observed in those integrals, especially 
for $F(t)$.
In such a case, we judge that ``coarse-graining`` (CG) is appropriate.
Note that $F(t)$ is a linear function of $t$
when the CG Rabi frequency $\Omega$ is constant.

\begin{figure}
\begin{center}
\includegraphics[scale=0.5]{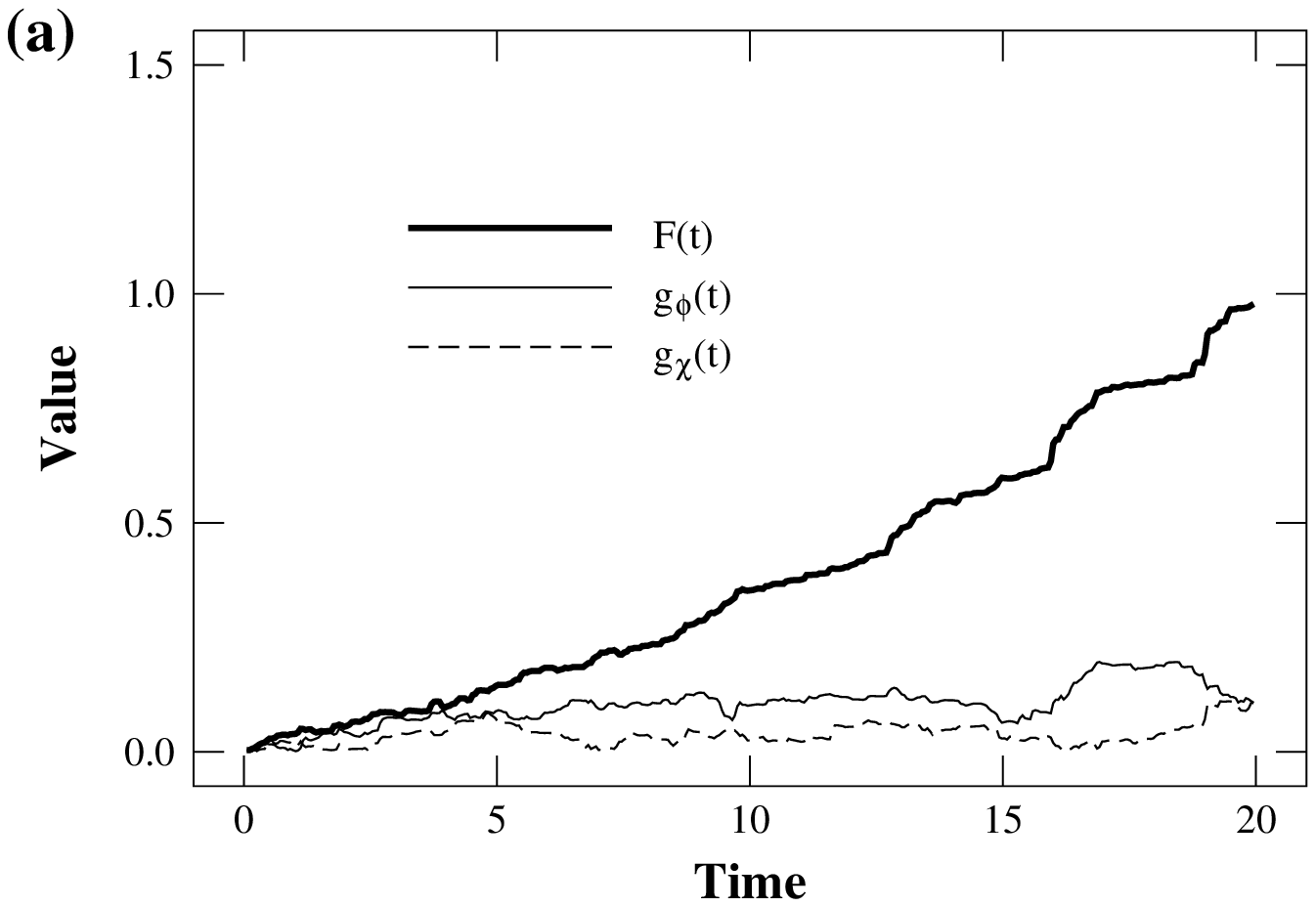}
\includegraphics[scale=0.5]{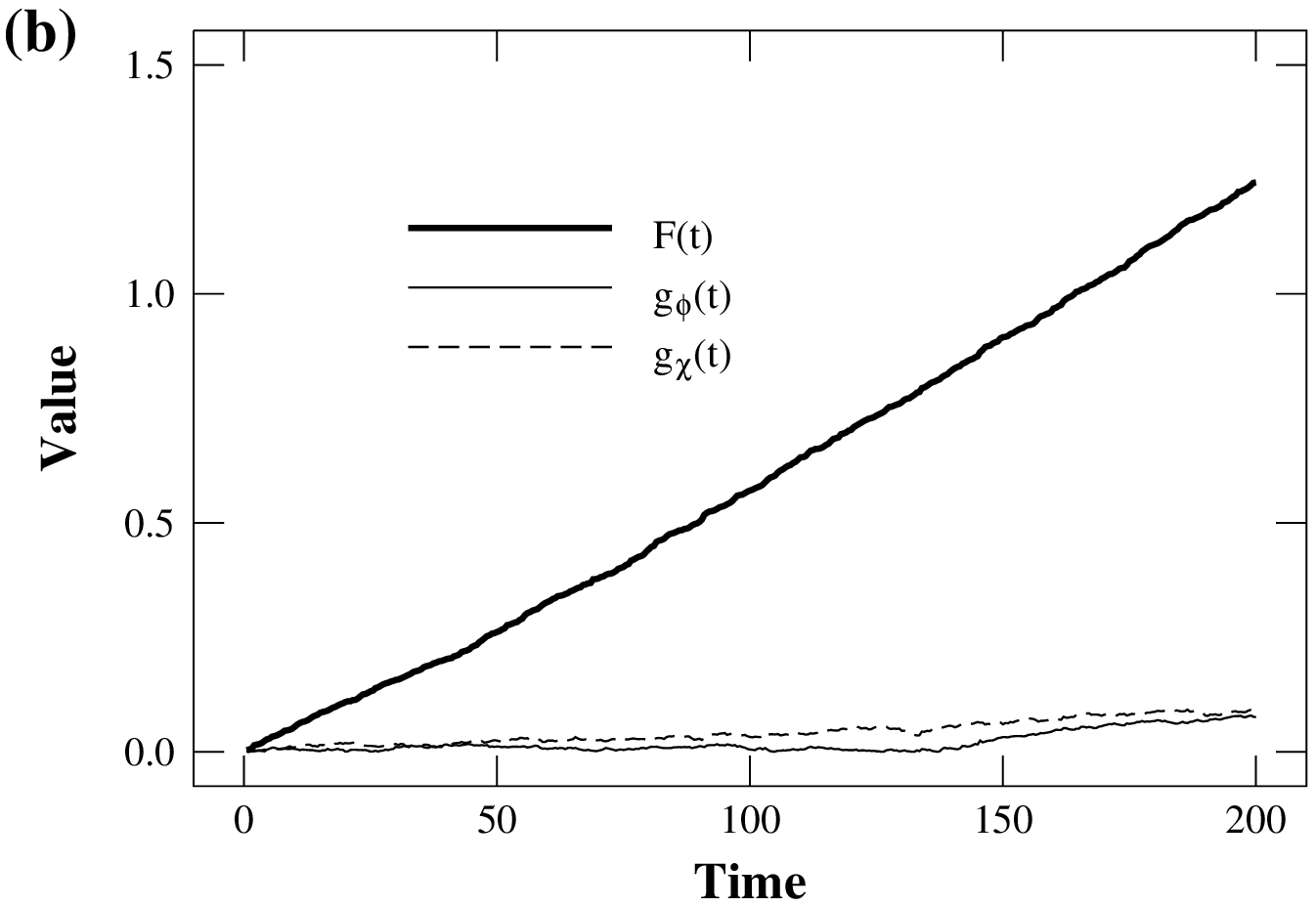}
\caption{
\label{fig:cond-rm}
Absolute values of the functions $F(t)$, $g_\phi(t)$, and $g_\chi(t)$
(see the main text) are shown: 
(a) $T=20$ and $\alpha=1$; (b) $T=200$ and $\alpha=10$.
The external fields used in these calculations are already shown
in Fig.~\ref{fig:rm020}(a) and Fig.~\ref{fig:rm200}(a), respectively.
}
\end{center}
\end{figure}

Figure \ref{fig:cond-rm} shows $|F(t)|$, $|g_\phi(t)|$, and $|g_\chi(t)|$
obtained from the numerical results in Sec.~\ref{sec:oct-random}.
For the case of $T=20$ in Fig.~\ref{fig:cond-rm}(a),
the values of $g_\phi(t)$ and $g_\chi(t)$ are small compared to $F(t)$,
but $F(t)$ cannot be considered as a linear function of $t$.
Thus, CG is not appropriate in this case.
On the other hand,
examining the case of $T=200$ in Fig.~\ref{fig:cond-rm}(b),
we realize that the condition, Eq.~(\ref{eqn:cond}),
is satisfied,
and $F(t)$ is regarded as a linear function of $t$.
Hence we conclude that CG for random matrix systems 
is appropriate for a rather large target time $T$,
and in such a case, the CG Rabi frequency becomes constant. 

\subsection{Smooth Transition between Random Vectors\label{sec:smooth}}

In Sec.~\ref{sec:oct-random},
we have already obtained 
the optimal field $\varepsilon(t)$ by the numerical calculation
for the random matrix systems, Eq.~(\ref{eqn:random-Hamiltonian}).
However,
only the overlap between the time-evolving controlled 
state $|\phi(t)\rangle$ and 
the target state $|\varphi_f\rangle$ was shown there.
In this section, we show the overlaps 
between the time-dependent states
defined by Eq.~(\ref{eqn:basis}) and $|\phi(t)\rangle$,
and find a smooth transition picture. 

\begin{figure}
\begin{center}
\includegraphics[scale=0.5]{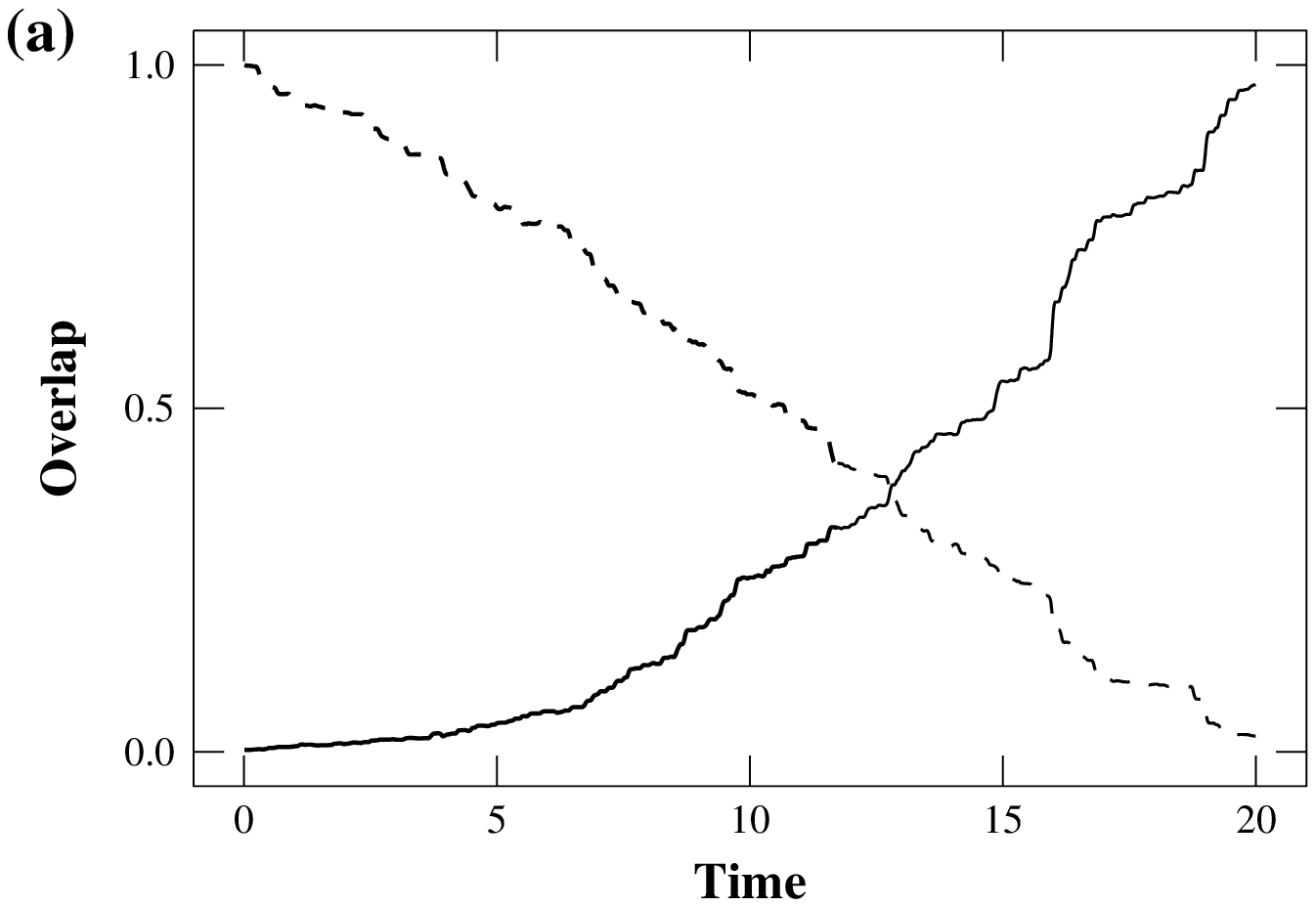}
\includegraphics[scale=0.5]{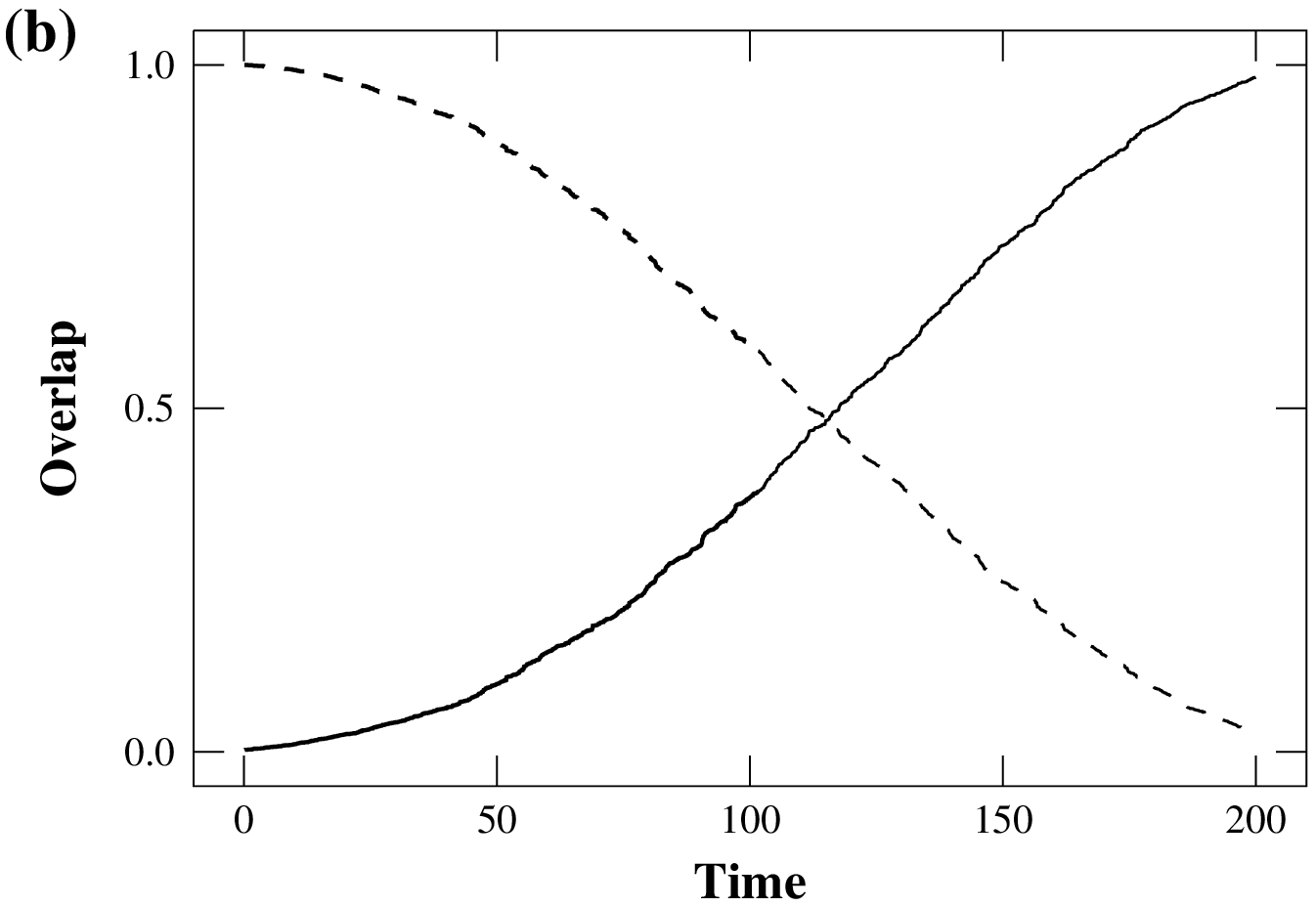}
\caption{
\label{fig:tr-rm}
The overlaps $|\langle\phi_0(t)|\phi(t)\rangle|^2$ and
$|\langle\chi_0(t)|\phi(t)\rangle|^2$ are shown:
(a) $T=20$ and $\alpha=1$; (b) $T=200$ and $\alpha=10$.
The external fields used in these calculations are already shown
in Fig.~\ref{fig:rm020}(a) and Fig.~\ref{fig:rm200}(a), respectively.
}
\end{center}
\end{figure}

In Fig.~\ref{fig:tr-rm}(a),
we show the overlap $|\langle\phi_0(t)|\phi(t)\rangle|^2$ and
$|\langle\chi_0(t)|\phi(t)\rangle|^2$ which are obtained
from the dynamics driven by the same external field shown
in Fig.~\ref{fig:rm020}(a).
Those curves in the figure are not smooth,
and it seems to be difficult to approximate them by 
the CG Rabi state, Eq.~(\ref{eqn:cg-rabi}), with a constant $\Omega$.
In Fig.~\ref{fig:tr-rm}(b),
on the other hand,
we see a smooth transition
from $|\phi_0(t)\rangle$ to $|\chi_0(t)\rangle$,
which is induced by the optimal field shown in Fig.~\ref{fig:rm200}(a).
In this case, the dynamics can be well represented by
the CG Rabi state with a constant $\Omega$.

\section{Analytic Expression for the Optimal Field\label{sec:analytic}}

In the previous sections, we have studied the controlled dynamics
when an optimal field is first given by the ZBR-OCT scheme.
In this section, in turn,
we first assume that the dynamics is well approximated
by the CG Rabi state,
and try to derive an analytic optimal field by using OCT \cite{TF04}.

\subsection{Coarse-Grained Transition Element}

We start from an assumption that
optimally controlled quantum states are represented by the CG Rabi states,
i.e., the forwardly evolving state $|\phi(t)\rangle$ and
the inversely evolving state $|\chi(t)\rangle$ are assumed to be
\begin{eqnarray}
\label{eqn:forward}
  |\phi(t)\rangle
  &=&|\phi_0(t)\rangle\cos|\Omega|t
   -ie^{-i\theta}|\chi_0(t)\rangle\sin|\Omega|t\\
  |\chi(t)\rangle
  &=&-ie^{i\theta}|\phi_0(t)\rangle\sin|\Omega|(t-T)\nonumber\\
   &&\qquad+|\chi_0(t)\rangle\cos|\Omega|(t-T).
\label{eqn:backward}
\end{eqnarray}
As we have seen numerically in Sec.~\ref{sec:smooth},
the optimal field induces a smooth transition
between $|\phi_0(t)\rangle$ and $|\chi_0(t)\rangle$.
In this section, we employ OCT to study
an analytic formulation of the optimal field.
Substituting Eqs.~(\ref{eqn:forward}) and (\ref{eqn:backward}) into
the expression of the optimal field, Eq.~(\ref{eqn:field}),
and after some manipulations, we obtain
\begin{equation}
\label{eqn:cg-field}
  \varepsilon(t)={\sin2|\Omega|T\over2\alpha\hbar}{\rm Re}\left[
    e^{-i\theta}\langle\phi_0(t)|V|\chi_0(t)\rangle
  \right],
\end{equation}
where $|\langle\phi_0(t)|\chi_0(t)\rangle| \ll 1$ has been used
as before.
This is an analytic expression for the optimal field while
the value of the CG Rabi frequency $\Omega$ and the phase parameter $\theta$
have not been determined yet.

The definition of the CG Rabi frequency, Eq.~(\ref{eqn:omega}), 
is used to determine $|\Omega|$.
Substituting Eq.~(\ref{eqn:cg-field}) and
using the relation $\Omega=e^{i\theta}|\Omega|$,
we obtain
\begin{equation}
\label{eqn:cg-omega}
  |\Omega|={\bar V^2\sin2|\Omega|T\over4\alpha\hbar^2}
\end{equation}
where
\begin{eqnarray}
\bar V^2&\equiv&\left<\left|\langle\phi_0(t)|V|\chi_0(t)\rangle\right|^2\right.
\nonumber\\
&&\qquad\left.+\left[e^{-i\theta}\langle\phi_0(t)|V|\chi_0(t)\rangle\right]^2
  \right>_{\rm CG}
\label{eqn:cg-v}
\end{eqnarray}
is a CG transition element.
This equation gives $|\Omega|$ when the penalty factor $\alpha$ and
the target time $T$ are fixed.
For a large $T$,
the second term in the right-hand side
is considered small compared to the first term.
In order to see this, we represent the initial and final state
using the eigenstates $|\phi_k\rangle$ of $H_0$ as 
\begin{equation}
  |\varphi_i\rangle=\sum_jc_j|\phi_j\rangle,\qquad
  |\varphi_f\rangle=\sum_kd_k|\phi_k\rangle,
\end{equation}
with the coefficients $c_j$ and $d_j$.
For a large $T$, we can ignore oscillating terms to obtain
\begin{eqnarray}
  &&\left|\langle\phi_0(t)|V|\chi_0(t)\rangle\right|^2\nonumber\\
  &&\qquad =\sum_{j,k}|c_j|^2|V_{jk}|^2|d_k|^2+\left|R(T)\right|^2\\
  &&\left[\langle\phi_0(t)|V|\chi_0(t)\rangle\right]^2
  =\left(R(T)\right)^2
\end{eqnarray}
where
\begin{equation}
\label{eqn:sum}
  R(T)\equiv\sum_jc_j^*V_{jj}d_je^{-E_jT/i\hbar},
\end{equation}
becomes small for $N\rightarrow\infty$
when $|\varphi_i\rangle$ and $|\varphi_f\rangle$ are
random vectors without any special correlation.
Thus Eq.~(\ref{eqn:cg-omega}) is simplified as 
\begin{equation}
  \bar V^2\approx\sum_{j,k}|c_j|^2|V_{jk}|^2|d_k|^2.
\end{equation}
If the following condition
\begin{equation}
\label{eqn:Tcond}
  {\bar V^2T\over2\alpha\hbar^2}>1,
\end{equation}
is satisfied, at least one $|\Omega|$ ($\Omega\ne0$) is
obtained from Eq.~(\ref{eqn:cg-omega}).
Using this $|\Omega|$, 
the final overlap $J_0$ is given by
\begin{equation}
  J_0=\sin^2|\Omega|T,
\label{eqn:j0}
\end{equation}
and the averaged amplitude $\bar\varepsilon$ of the external field
(\ref{eqn:cg-field}) is calculated as 
\begin{equation}
  \bar\varepsilon\equiv
  \sqrt{{1\over T}\int_0^T|\varepsilon(t)|^2dt}
  \approx{\sqrt2\hbar|\Omega|\over\bar V}.
\label{eqn:eave}
\end{equation}

\begin{figure}
\begin{center}
\includegraphics[scale=0.5]{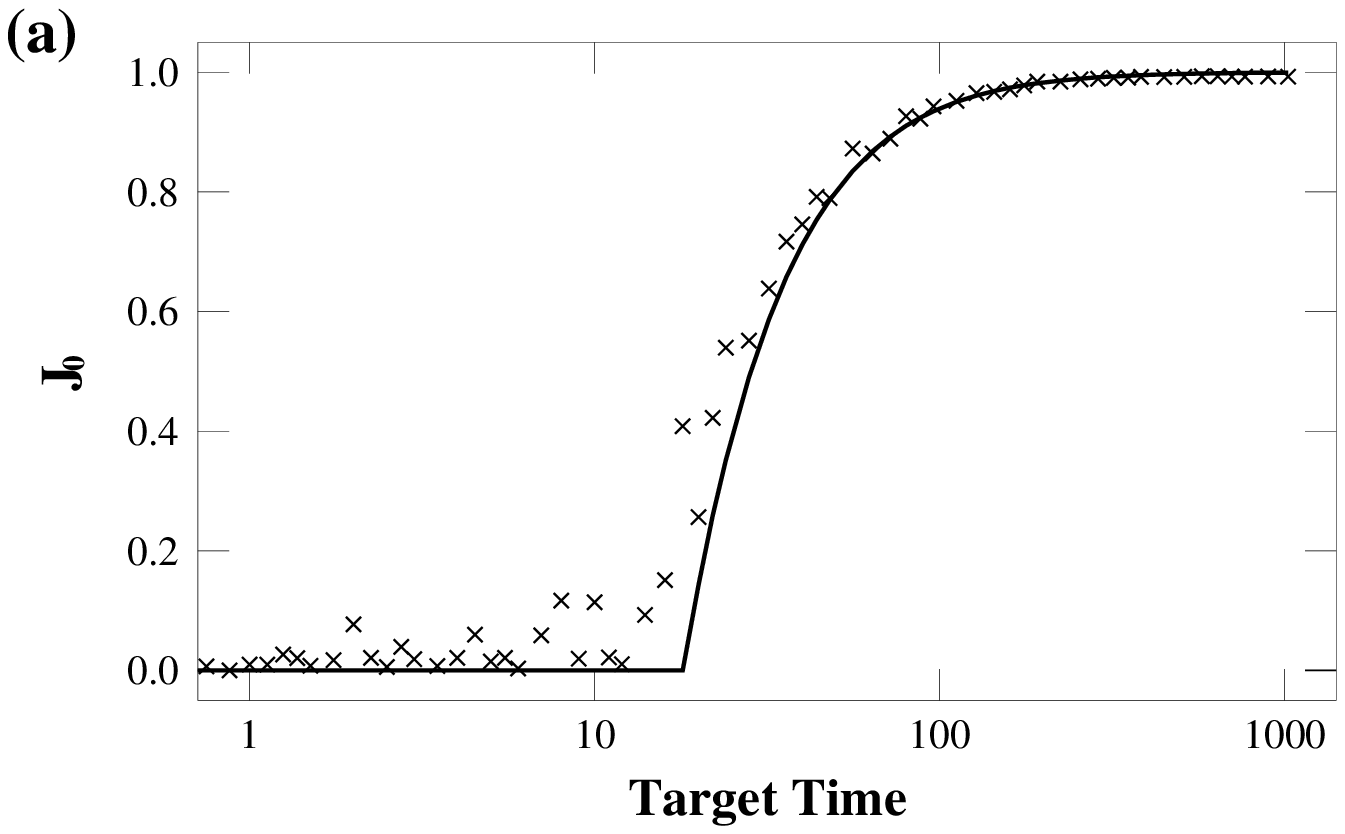}
\includegraphics[scale=0.5]{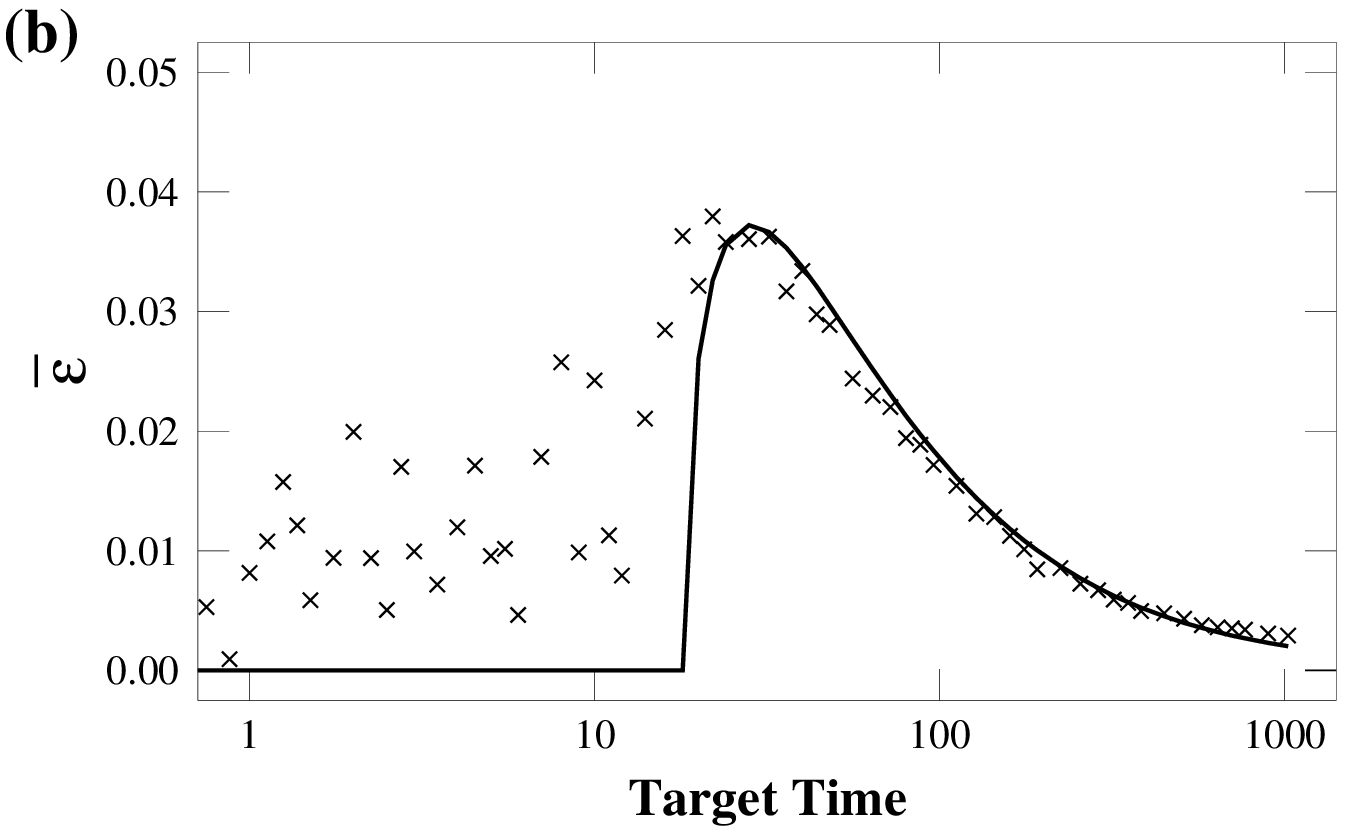}
\caption{
\label{fig:cg-values}
(a) The final overlap
$J_0=\left|\langle\phi(T)|\varphi_f\rangle\right|^2$ and
(b) the averaged field amplitude $\bar\varepsilon$
for a $64\times64$ random matrix system are shown as a
function of the target time $T$.
Marks ($\times$) represent the numerical results by the Zhu-Botina-Rabitz scheme.
Solid curves represent our analytic results under the assumption of 
the CG Rabi state. 
}
\end{center}
\end{figure}

In Fig. \ref{fig:cg-values}, we compare the predicted values, 
Eqs.~(\ref{eqn:j0}) and (\ref{eqn:eave}),
with the numerical results for the random matrix system.
Those results agree well each other especially for a large $T$, 
i.e., the CG picture is valid and useful especially for a large 
target time $T$.\footnote{
Note that
there exists a threshold $T_c\equiv2\alpha\hbar^2/\bar V^2$,
the smallest target time satisfying the condition, Eq.~(\ref{eqn:Tcond}).
If we choose a smaller $T$ than $T_c$,
there is no external field which induces the smooth transition
described by the CG Rabi state.
On the other hand, the numerical method can give finite solutions 
for such cases because there is no assumption (restriction) 
about the dynamics except that it obeys to the Schr\"odinger equation.}

\subsection{Analytic Solution for Perfect Control}

\begin{figure}
\begin{center}
\includegraphics[scale=0.5]{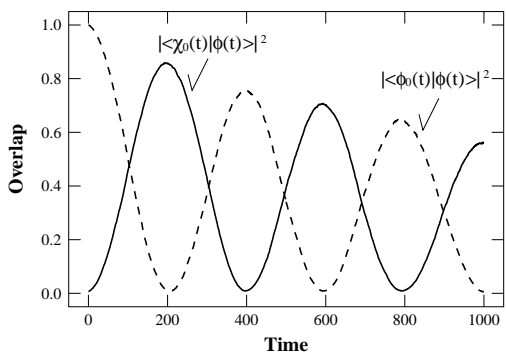}
\caption{
\label{fig:rabi}
The coarse-grained Rabi oscillation induced by the analytical
external field for perfect control is shown
for the case $k=3$ in Eq. (\ref{eqn:perfect-field}).
The solid curve represents $|\langle\chi_0(t)|\phi(t)\rangle|^2$,
and the dashed $|\langle\phi_0(t)|\phi(t)\rangle|^2$.
The initial and the target states are Gaussian random vectors
in a $256\times256$ GOE random matrix Hamiltonian system.
}
\end{center}
\end{figure}

\begin{figure}
\begin{center}
\includegraphics[scale=0.6]{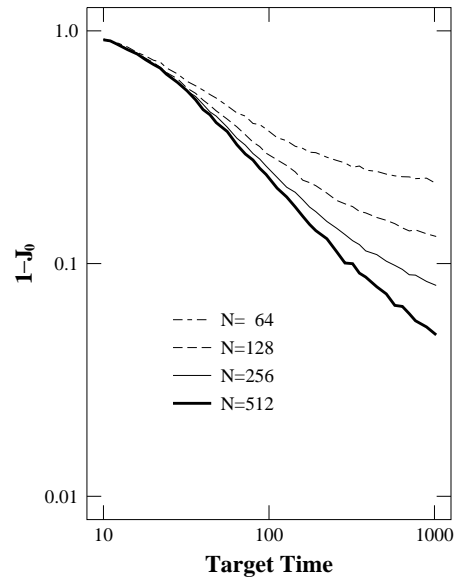}
\caption{
\label{fig:pfct}
The target-time dependence of the final overlap $J_0$ by
the analytic optimal field with $k=1$ is shown.
The residual probability $1-J_0$ from the perfect control $J_0=1$ is
depicted for various matrix sizes $N$ of GOE random Hamiltonians.
The initial and the final states are Gaussian random vectors.
}
\end{center}
\end{figure}

In the ZBR scheme,
we must choose a small penalty factor $\alpha$
to make the final overlap large enough.
In our analytical results,
if we take the limit $\alpha\rightarrow0$, 
we find that 
\begin{equation}
|\Omega|=\frac{(2k-1)\pi}{2T} \quad (k=1,2,...)
\end{equation}
satisfies Eq.~(\ref{eqn:cg-omega}),
and then $J_0=1$, i.e., a perfect control
is achieved. 
Using Eqs.~(\ref{eqn:cg-field}) and (\ref{eqn:cg-omega}),
the optimal field for the perfect control in the small $\alpha$ 
limit is obtained as 
\begin{equation}
\label{eqn:perfect-field}
  \varepsilon(t)={(2k-1)\pi\hbar\over\bar V^2T}{\rm Re}\left[
     e^{-i\theta}\langle\phi_0(t)|V|\chi_0(t)\rangle
  \right]
\end{equation}
where $\theta$ can be determined by a normalization condition as
\begin{equation}
e^{2 i \theta} = \frac{\langle \phi_0(T) | \varphi_f \rangle}
{\langle \varphi_f| \phi_0(T) \rangle}.
\end{equation}
This field is expected to be the optimal field
which steers the quantum state $|\varphi_i\rangle$ at $t=0$
to $|\varphi_f\rangle$ at $t=T$,
as well as it induces a CG Rabi oscillation
between $|\phi_0(t)\rangle$ and $|\chi_0(t)\rangle$.
Note that the penalty factor $\alpha$ does not appear
in Eq.~(\ref{eqn:perfect-field}),
so this is different from other non-iterative 
optimal fields discussed in \cite{ZR99}.

We next examine when and how the analytic optimal 
field works for a random matrix system ($256\times256$ GOE random matrix).
Figure \ref{fig:rabi} demonstrates the coarse-grained Rabi oscillation
induced by the analytic field, Eq.~(\ref{eqn:perfect-field}), with $k=3$,
where smooth oscillations of $|\langle\phi_0(t)|\phi(t)\rangle|^2$
and $|\langle\chi_0(t)|\phi(t)\rangle|^2$ are observed.
The initial and the target states are both Gaussian random vectors
with $256$ elements.
This result shows that the field actually
produces the CG Rabi oscillation in the random matrix system.

Finally, in Fig.~\ref{fig:pfct},
we show the performance of the analytic field, Eq.~(\ref{eqn:perfect-field}),
for the same type of control problem with various matrix sizes.
The abscissa and the ordinate are the target time $T$ and
the residual probability $1-J_0$, respectively.
This result shows that the final overlap $J_0$ approaches unity,
i.e., the perfect control is achieved as 
the target time and the matrix size become both large.

\section{Summary and Discussion}

We have studied optimal control of random matrix systems and a quantum kicked rotor
as examples of quantum chaos systems.
Using the ZBR-OCT scheme, we numerically achieved almost perfect control
for the above systems where the initial state $|\varphi_i \rangle$ 
and the target state $|\varphi_f \rangle$ 
are random vectors (except the case of a quantum kicked rotor with $K=1$). 
However, the optimal fields and the overlap 
$|\langle \phi(t) |\varphi_f \rangle|^2$ thus obtained 
are too complicated to be analyzed as shown in Figs.~\ref{fig:rm020}, 
\ref{fig:rm200}, \ref{fig:fdK1}, and \ref{fig:fdK7}.
On the other hand, as shown in Fig.~\ref{fig:tr-rm},
the overlaps $|\langle \phi_0(t) |\phi(t) \rangle|^2$ 
and $|\langle \chi_0(t) |\phi(t) \rangle|^2$ 
are rather smooth where $|\phi_0(t)\rangle$ 
($|\chi_0(t)\rangle$) 
represents a free forward (backward) evolution of the system, 
so we can introduce coarse 
grained concepts: a CG Rabi state and a CG Rabi frequency.
The CG Rabi state is an analogue of a usual Rabi state but 
it describes a transition between $|\phi_0(t)\rangle$ and $|\chi_0(t)\rangle$
as in Eq.~(\ref{eqn:cg-rabi}).
The CG Rabi frequency is defined
by ignoring rapidly oscillating terms as in Eq.~(\ref{eqn:omega}).
We applied this picture to OCT and 
obtained an analytic expression for the optimal field, Eq.~(\ref{eqn:perfect-field}).
We also numerically confirmed that the analytic field 
actually works in controlling random vectors
when the target time and the matrix size are both large enough.
\\

In closing, we discuss future directions of this study:
(a) We mainly studied strong-chaos limit cases as described by 
random matrix Hamiltonians, and applied the coarse grained ideas to them.
Thus, the next problem should be addressed on less chaotic cases 
as described by banded random matrix Hamiltonians.
A quantum kicked rotor with a small $K$ 
will be a good example for that purpose \cite{GB01}.
(b) The other interesting problem is 
the semiclassical limit of the controlled dynamics.
Though we have shown that quantum chaos systems 
can be controlled, we don't know its semiclassical 
behavior since there are many difficulties
in taking the semiclassical limit $\hbar\rightarrow0$.
There are, on the other hand, many works
studying {\it chaos control} in classical mechanics,
and there are some examples utilizing stochastic features
of phase space in ``targeting'' problems \cite{Ott02,SR91}.
In this respect,
it is strongly desirable to study chaos control
from semiclassical points of view \cite{BB02,semiclassical}.
(c) In connection with quantum information processings,
control of quantum entanglement in quantum chaos systems \cite{entanglement}
will be another interesting subject to be pursued.  

\begin{acknowledgments}
The authors thank Prof.~S.~A.~Rice, Prof.~H.~Rabitz, Prof.~M.~Toda,
Prof.~H.~Nakamura, Prof.~H.~Kono, Prof.~S.~Tasaki, 
Prof.~A.~Shudo, Dr.~Y.~Ohtsuki, and Dr.~G.~V.~Mil'nikov
for useful discussions.
\end{acknowledgments}

%\end{document}

\end{document}